\documentclass{easychair}

\title{Logic-Induced Bisimulations}
\author{
       Jim de Groot\inst{1} 
  \and Helle Hvid Hansen\inst{2} 
  \and Alexander Kurz\inst{3}}
\institute{
       The Australian National University, Canberra, Australia \\
       \email{jim.degroot@anu.edu.au}
  \and University of Groningen, Groningen, The Netherlands \\
       \email{h.h.hansen@rug.nl}
  \and Chapman University, Orange, California, USA \\
       \email{akurz@chapman.edu}}
\date{}

\authorrunning{}
\titlerunning{}

\usepackage{amsmath}
\usepackage{amssymb}
\usepackage{amsthm}
  \theoremstyle{definition}
  \swapnumbers
    \newtheorem{para}{}[section]
    \newtheorem{definition}[para]{Definition}
    \newtheorem{example}[para]{Example}
    
    \newtheorem{remark}[para]{Remark}
    
  \theoremstyle{theorem}
    \newtheorem{lemma}[para]{Lemma}
    \newtheorem{corollary}[para]{Corollary}
    \newtheorem{theorem}[para]{Theorem}
    \newtheorem{proposition}[para]{Proposition}

\usepackage{stmaryrd}
\usepackage{tikz-cd}
  \usetikzlibrary{arrows,arrows.meta}
  \tikzcdset{arrow style=tikz, diagrams={>=stealth'}}
  \newcommand*\circled[1]{{\footnotesize \tikz[baseline=(char.base)]{
            \node[shape=circle,draw,inner sep=2pt] (char) {#1};}}}
  
\usepackage[mathscr]{euscript}
\usepackage{stackengine}

\newcommand{\ms}[1]{\mathscr{#1}}

\newcommand{\cat}[1]{\mathbf{#1}}
\newcommand{\fun}[1]{\mathsf{#1}}

  {\begin{color}{blue}}{\end{color}}

\renewcommand{\hat}[1]{\widehat{#1}}
\renewcommand{\tilde}[1]{\widetilde{#1}}
\newcommand{\ov}[1]{\overline{#1}}
\renewcommand{\phi}{\varphi}
\newcommand{\und}[1]{\underline{#1}}
\renewcommand{\epsilon}{\varepsilon}

\DeclareMathOperator{\id}{id}
\DeclareMathOperator{\Id}{Id}

\DeclareMathOperator{\Hom}{Hom}

\renewcommand{\th}{\operatorname{th}}
\newcommand{\sem}[1]{\llb #1 \rrb}
\newcommand{\initobj}{\mathbf{0}} 
\mathchardef\hyphen="2D
\newcommand{\trans}[1]{#1^{\sharp}}
\newcommand{\mate}[1]{#1^{\flat}}

\newcommand{\unitc}{\eta^{\cat{C}}}
\newcommand{\unita}{\eta^{\cat{A}}}

\newcommand{\llb}{\llbracket}
\newcommand{\rrb}{\rrbracket}
\newcommand{\llangle}{\langle\!\langle}
\newcommand{\rrangle}{\rangle\!\rangle}

\renewcommand{\Box}{\boxempty}
\newcommand{\dbox}{\boxbar}
\newcommand{\ddiamond}{\mathbin{\rotatebox[origin=c]{45}{$\boxslash$}}}

\newcommand{\la}{\leftarrow}
\newcommand{\too}{\longrightarrow}
\newcommand{\injr}{\hookrightarrow}

\newcommand{\dpo}[1]{\ov{#1}} 
\newcommand{\po}[1]{\widehat{#1}} 

\newcommand{\Pow}{\mathcal{P}} 
\newcommand{\Bspan}{(B,\pi_1,\pi_2)}
\newcommand{\posRel}{\cat{Rel}(X_1,X_2)}
\newcommand{\CoalgT}{\cat{Coalg}(\fun{T})}
\newcommand{\RegEpiMono}{(\ms{R}eg\ms{E}pi, \ms{M}ono)}
\newcommand{\cPow}{\mathcal{Q}} 
\newcommand{\Ltr}{\fun{L}^{\mathsf{tr}}}
\newcommand{\Lhm}{\fun{L}^{\mathsf{hm}}}

\newcommand{\rhotr}{\rho^{\mathsf{tr}}}
\newcommand{\rhohm}{\rho^{\mathsf{hm}}}
\newcommand{\Pba}{\cPow_{\mathsf{BA}}}
\newcommand{\Uf}{\fun{Uf}}

\newcommand{\diam}[1]{\langle #1 \rangle}

\begin{document}

\maketitle

  \begin{abstract}
  We define a new logic-induced notion of bisimulation (called $\rho$-bisimulation) 
  for coalgebraic modal logics given by a logical connection, and investigate its 
  properties. We show that it is structural in the sense that it is defined only
  in terms of the coalgebra structure and the one-step modal semantics and, moreover, 
  can be characterised by a form of relation lifting.
  Furthermore we compare $\rho$-bisimulations to several well-known equivalence notions, 
  and we prove that the collection of bisimulations between two models often forms a 
  complete lattice.
  The main technical result is a Hennessy-Milner type theorem which states that, 
  under certain conditions, logical equivalence implies $\rho$-bisimilarity. 
  In particular, the latter does \emph{not} rely on a duality between functors 
  $\fun{T}$ (the type of the coalgebras) and $\fun{L}$ (which gives the logic),
  nor on properties of the logical connection $\rho$.
  \end{abstract}
%

\section{Introduction}\label{sec:intro}

  In this paper, we investigate when logical equivalence for a given modal
  language can be captured by a structural semantic equivalence notion,
  understood as a form of bisimulation.
  Our investigation is carried out in the setting of coalgebraic modal logic
  \cite{KupPat11}, where semantic structures are given by coalgebras
  for a functor $\fun{T} \colon \cat{C} \to \cat{C}$ \cite{Rut00}.
  This allows for a uniform treatment of a wide variety of modal
  logics~\cite{KupPat11,Pat03b,Sch08}.
  Coalgebras come with general notions of \emph{behavioural equivalence} 
  and \emph{bisimilarity}, and a logic is said to be \emph{expressive} if 
  logical equivalence implies behavioural equivalence,
  in which case we have a generalisation of the classic Hennessy-Milner
  theorem \cite{HenMil85}.

  For $\cat{Set}$-coalgebras, i.e., when $\cat{C} = \cat{Set}$,
  it has been shown that a coalgebraic modal logic is expressive if
  the language has sufficiently large conjunctions and 
  the set $\Lambda$ of modalities is \emph{separating},
  meaning that they separate points in $\fun T X$ \cite{Mos99,Pat04,Sch08}.
  In the more abstract setting of coalgebraic modal logic,
  where a logic is given by a functor  and its  semantics by a natural
  transformation $\rho$ \cite{BonKur05,Kli07},
  a sufficient condition for a logic being expressive is that the so-called
  mate of $\rho$ is pointwise monic \cite[Theorem 4.2]{Kli07}. 

  In this line of research,
  modal logics are often viewed as specification languages for coalgebras.
  Therefore behavioural equivalence is a given, and the aim is to find expressive logics.
  However, sometimes the modal language is of primary interest \cite{BalCin18} 
  and the relevant modalities need not be separating, see e.g.~\cite{FanWanDit14,BakDitHan17}.
  This leads us to consider the following question: 
  \begin{center}
    Given a possibly non-expressive coalgebraic modal logic, can we \\
    characterise logical equivalence by a notion of bisimulation?
  \end{center}
  Such investigations have been carried out earlier in \cite{BakHan17}
  where the notion of $\Lambda$-bisimulation was proposed for $\cat{Set}$-coalgebras
  and coalgebraic modal logics with a classical propositional base.

  Here we generalise and extend the work of \cite{BakHan17} beyond $\cat{Set}$
  using the formulation of coalgebraic modal logic via dual adjunctions 
  \cite{BonKur05,Kli07,KurRos12}.
  Examples include coalgebras over ordered and topological spaces and
  modal logics on different propositional bases.
  After recalling basic definitions of coalgebraic modal logic in Section~\ref{sec:cml-rel},
  we define the concept of a \emph{$\rho$-bisimulation} in Section~\ref{sec:rho-bis}.
  For $\cat{Set}$-coalgebras, this is a relation $B$ between coalgebras for which the 
  so-called $B$-coherent pairs \cite{HanKupPac09,BalCin18} give rise to a congruence
  between complex algebras. 

  The definition of $\rho$-bisimulation is structural in the sense that it is
  defined in terms of the coalgebra structure and the one-step modal semantics $\rho$.
  Moreover, it can often be characterised as a greatest fixpoint via relation lifting.
  For coalgebras on finite sets, this means that $\rho$-bisimilarity can be
  computed by a partition refinement algorithm.
  We also prove results concerning truth-preservation,
  composition and lattice structure.

  The main technical results are found in Section~\ref{sec:disting}
  and concern the distinguishing power of {$\rho$-bisimulations}.
  We first compare $\rho$-bisimulations with other coalgebraic equivalence notions.
  Subsequently, we prove a Hennessy-Milner style theorem (Theorem~\ref{thm:hm})
  in which we give conditions that guarantee that logical equivalence {is a} $\rho$-bisimulation.
  We emphasise that the logic is \emph{not} assumed to be expressive and
  $\rho$-bisimilarity will generally  differ from bisimilarity for $\fun{T}$-coalgebras.
  Finally, we define a notion of translation between logics and show that if
  the language of $\rho'$ is a propositional extension of the language of $\rho$,
  then $\rho$-bisimulations are also $\rho'$-bisimulations (Proposition~\ref{prop:tau-flat}). 

  By instantiating Proposition~\ref{prop:tau-flat},
  we obtain that for labelled transition systems the $\rho$-bisimilarity notions
  for Hennessy-Milner logic \cite{HenMil85} and trace logic \cite{Kli07} coincide
  and are equal to the standard notion of bisimilarity even without assuming image-finiteness. 
  These two logics have the same modalities, which are separating,
  but trace logic has $\top$ as the only propositional connective.

\paragraph{Earlier version}
  This is the extended version of an AIML paper \cite{GroHanKur20}
  with the same name.
  The current paper includes proofs that were left out in \cite{GroHanKur20}.
  Besides, it includes an additional example of a logic for linear weighted 
  automata that matches precisely the logic from \cite[Section 3.2]{BonEA12}
  (Example \ref{exm:trace-vec}),
  and a Hennessy-Milner result for it (Example \ref{exm:trans-vec}).

\section{Coalgebraic modal logic}\label{sec:cml-rel}

  We review some background on coalgebraic logic, categorical algebra, and Stone duality.
  For more details, e.g.~\cite{Rut00,KupPat11,AdaHerStr90,ARV,Joh82}.
  We write $\cat{Set}$ for the category of sets and functions.

  Coalgebraic modal logic generalises modal logic from Kripke frames to
  coalgebras for a functor $\fun{T}$.

\medskip\noindent\textbf{Coalgebras} can be understood as generalised,
  state-based systems defined parametrically in the system type $\fun{T}$.
  Formally, we require $\fun{T}$ to be an endofunctor on a category $\cat{C}$.
  A $\fun{T}$-\emph{coalgebra} is then a pair $(X, \gamma)$
  such that $\gamma : X \to \fun{T}X$ is a morphism in $\cat{C}$.
  The object $X$ is the state space, and the arrow $\gamma$
  is the coalgebra structure map.
  A \emph{$\fun{T}$-coalgebra morphism} from $(X, \gamma)$ to
  $(X', \gamma')$ is a $\cat{C}$-morphism $f : X \to X'$ satisfying
  $\gamma' \circ f = \fun{T}f \circ \gamma$.
  Together, $\fun{T}$-coalgebras and $\fun{T}$-coalgebra morphisms form a category
  which we write as $\cat{Coalg}(\fun{T})$.

\medskip\noindent An \textbf{algebra} for a functor is the dual notion of a coalgebra.
  Given an endofunctor $\fun{L} \colon \cat{A} \to \cat{A}$,
  an \emph{$\fun{L}$-algebra} is a pair $(A, \alpha)$ such that $\alpha : \fun{L}A \to A$
  is a morphism in $\cat{A}$. An \emph{$\fun{L}$-algebra morphism} from $(A,\alpha)$ to
  $(A',\alpha')$ is an $\cat{A}$-morphism $h: A \to A'$ such that
  $h \circ \alpha = \alpha' \circ \fun{L}h$. We write $\cat{Alg}(\fun{L})$
  for the category of $\fun{L}$-algebras and $\fun{L}$-algebra morphisms.

\begin{example}\label{ex:Kripke-coalg}
  A Kripke frame $(X, R\subseteq X \times X)$ is a coalgebra for the
  covariant powerset functor $\Pow\colon \cat{Set}\to\cat{Set}$ which maps
  a set to its set of subsets, and a function $f\colon X \to Y$ to the 
  direct image map $f[-] \colon \Pow X \to \Pow Y$,
  by defining $\gamma \colon X \to \Pow X$ as $\gamma(x) = R[x] = \{ y \in X \mid xRy \}$.
  Similarly, a Kripke model $(X, R, V)$, where $V$ is a valuation of a set $P_0$
  of atomic propositions, is a coalgebra for the $\cat{Set}$-functor
  $\Pow(-) \times \Pow(P_0)$ (which is constant in its second component)
  by taking $\gamma(x) = (R[x],V'(x))$, with $V'(x) = \{ p \in P_0 \mid x \in V(p)\}$.
  It can be verified that the ensuing notion of coalgebra morphism
  coincides with the usual notion of bounded morphism for Kripke frames and Kripke models,
  respectively.
\end{example}

\begin{example}\label{ex:LTS-coalg}
  Labelled transition systems (LTSs) are coalgebras for
  the $\cat{Set}$-functor $\fun{T} = \Pow(-)^A$
  where $\Pow$ is the covariant powerset functor and $A$ is the set of labels.
  A coalgebra $\gamma \colon X \to \Pow(X)^A$ 
  specifies for each state $x \in X$ and label $a \in A$,
  the set $\gamma(x)(a)$ of $a$-successors of $x$.
  In other words, an LTS is an $A$-indexed multi-relational Kripke frame.
  One readily verifies that coalgebra morphisms are $A$-indexed bounded morphisms.  
\end{example}

\paragraph{Logical connections}
  To investigate logics for $\fun{T}$-coalgebras in this generality,
  we use the Stone duality approach to modal logic \cite{Gol76,Abr91},
  but rather than a full duality, here one requires only a dual adjunction
  $
  \begin{tikzcd}[sep=1.5em, cramped]
    \fun{P} : \cat{C}
          \arrow[r, shift left=1.7pt]
      & \cat{A} : \fun{S}
          \arrow[l, shift left=1.7pt]
  \end{tikzcd}
  $
  (sometimes called a \emph{logical connection})
  between a category $\cat{C}$ of state spaces
  and a category $\cat{A}$ of algebras that encode a propositional base logic.
  We emphasise that the functors $\fun{P}$ and $\fun{S}$ are contravariant.
  The classic example is then the instance 
  $
  \begin{tikzcd}[sep=1.5em, cramped]
    \Pba : \cat{Set}
          \arrow[r, shift left=1.7pt]
      & \cat{BA} : \Uf
          \arrow[l, shift left=1.7pt]
  \end{tikzcd}
  $
  where $\Pba$ maps a set to its Boolean algebra of predicates (i.e., subsets),
  and $\Uf$ maps a Boolean algebra to its set of ultrafilters. 

  We denote the units of a dual adjunction
  $
  \begin{tikzcd}[sep=1.5em, cramped]
    \fun{P} : \cat{C}
          \arrow[r, shift left=1.7pt]
      & \cat{A} : \fun{S}
          \arrow[l, shift left=1.7pt]
  \end{tikzcd}
  $
  by $\unitc\colon \Id_{\cat{C}} \to \fun{SP}$ and
  $\unita \colon \Id_{\cat{A}} \to \fun{PS}$,
  and the bijection of Hom-sets $\cat{C}(C,\fun{S}A) \cong \cat{A}(A,\fun{P}C)$
  in both directions by $\trans{(-)}$.
  Recall that for $f \colon C \to \fun{S}A$, the adjoint transpose of $f$
  is $\trans{f} = \fun{P}f \circ \unita_A$, and for $g\colon A \to \fun{P}C$,
  the adjoint is $\trans{g} = \fun{S}g \circ \unitc_C$.

\paragraph{Coalgebraic Modal Logic}    
  Given a dual adjunction
  $
  \begin{tikzcd}[sep=1.5em, cramped]
    \fun{P} : \cat{C}
          \arrow[r, shift left=1.7pt]
      & \cat{A} : \fun{S}
          \arrow[l, shift left=1.7pt]
  \end{tikzcd}
  $
  and an endofunctor $\fun{T}$ on $\cat{C}$,
  a \emph{modal logic for $\fun{T}$-coalgebras}
  is a pair $(\fun{L}, \rho)$ consisting of an endofunctor
  $\fun{L}\colon \cat{A}\to \cat{A}$ (defining modalities)
  and a natural transformation $\rho : \fun{LP} \to \fun{PT}$,
  (defining the \emph{one-step modal semantics}).
  This data gives rise to a functor
  $\cat{Coalg}(\fun{T}) \to \cat{Alg}(\fun{L})$ which sends a coalgebra
  $(X, \gamma)$ to its \emph{complex algebra} $(\fun{P}X, \gamma^*)$,
  where $\gamma^* = \fun{P}\gamma \circ \rho_X$.
  Assuming that $\cat{Alg}(\fun{L})$ has an initial algebra
  $\alpha : \fun{L}\Phi \to \Phi$, which generalises the Lindenbaum-Tarski algebra,
  the semantics of (equivalence classes of) formulae is obtained
  as the unique $\cat{Alg}(\fun{L})$-morphism
  $\sem{-}_\gamma \colon (\Phi,\alpha) \to (\fun{P}X, \gamma^*)$.
  Viewing the semantics as an $\cat{A}$-morphism  $\sem{-}_\gamma \colon \Phi \to \fun{P}X$,
  its adjoint
  $\th_{\gamma} = \trans{\sem{-}_{\gamma}} \colon X \to \fun{S}\Phi$,
  is called the \emph{theory map}, since in the classic case
  it maps a state in $X$ to the ultrafilter of $\fun{L}$-formulae it satisfies.
  By their definitions, the semantics and the theory map make the following diagrams commute: 
  \[
    \begin{tikzcd}[row sep=1.4em, cramped]
      \fun{L}\Phi
            \arrow[rr, "\alpha"]
            \arrow[d, "\fun{L}\llb \cdot \rrb_{\gamma}" left]
        &
        & \Phi 
            \arrow[d, "\llb \cdot \rrb_{\gamma}"]
        & X \arrow[rr, "\th_{\gamma}"]
            \arrow[d, "\gamma"]
        &
        & \fun{S}\Phi
            \arrow[d, "\fun{S}\alpha"] \\
      \fun{LP}X
            \arrow[r, "\rho_X"]
        & \fun{PT}X
            \arrow[r, "\fun{P}\gamma"]
        & \fun{P}X
        & \fun{T}X
            \arrow[r, "\fun{T}\th_{\gamma}"]
        & \fun{TS}\Phi
            \arrow[r, "\rho_{\Phi}^{\flat}"] 
        & \fun{SL}\Phi
    \end{tikzcd}
  \]
Here $\mate{\rho}\colon \fun{TS} \to \fun{SL}$ is the so-called mate of $\rho$.
This is the natural transformation obtained (component-wise) as the adjoint of $\rho_{\fun{S}} \circ \fun{L} \unita$.

\begin{example}\label{ex:LTS-trace-logic}
Consider the self-dual adjunction
  $
  \begin{tikzcd}[sep=1.5em, cramped]
    \cPow : \cat{Set}
          \arrow[r, shift left=1.7pt]
      & \cat{Set} : \cPow
          \arrow[l, shift left=1.7pt]
  \end{tikzcd}
  $
given in both directions by the contravariant powerset functor $\cPow$,
which maps a set to its powerset $2^X$, and a function $f\colon X \to Y$
to its inverse image map $f^{-1}\colon 2^Y \to 2^X$.
In this case, the adjoints are given by transposing.
That is, for $f \colon X \to 2^Y$,  $\trans{f}\colon Y \to 2^X$ is defined by
$\trans{f}(y)(x) = f(x)(y)$.

Considering LTSs as $\Pow(-)^A$-coalgebras over $\cat{Set}$ (cf.~Example~\ref{ex:LTS-coalg}),
we obtain \emph{trace logic for LTSs} \cite[Example~3.2]{Kli07}
by taking $\Ltr \colon \cat{Set}\to\cat{Set}$ to be the functor
$\Ltr = 1+A\times(-)$ (where $1 = \{*\}$ is a set with one element).
This encodes a modal signature with a constant modality $\top$
and a unary modality for each $a \in A$.
Since $\cat{A} = \cat{Set}$, trace logic has no other connectives.
The initial $\Ltr$-algebra consists of finite sequences over $A$
with the empty word as constant, and prefixing with elements from $A$
as the unary operations. That is, $\Ltr$-formulae are of the form
$\diam{a_1} \cdots \diam{a_k}\top$, where $k \geq 0$.

We obtain the usual semantics of $\top$ and $A$-labelled diamonds
by defining the modal semantics $\rhotr\colon 1+A\times\cPow(-) \to \cPow(\Pow(-)^A)$ as
$\rhotr_X(*) = \Pow(X)^A$ and 
$\rhotr_X(a,U) = \{ t \in \Pow(X)^A \mid t(a) \cap U \neq \emptyset\}$.
Hence for an LTS $(X,\gamma)$,
$\sem{\diam{a_1} \cdots \diam{a_k}\top}_\gamma$ is the subset of $X$ consisting of
states $x$ that can execute the trace $a_1 \cdots a_k$.
\end{example}

\begin{example}\label{ex:LTS-HM-logic}
  Again consider LTSs as $\Pow(-)^A$-coalgebras over $\cat{Set}$
  (cf.~Example~\ref{ex:LTS-coalg}), but now take the classic dual adjunction
  $
  \begin{tikzcd}[sep=1.5em, cramped]
    \Pba : \cat{Set}
          \arrow[r, shift left=1.7pt]
      & \cat{BA} : \Uf
          \arrow[l, shift left=1.7pt]
  \end{tikzcd}
  $.
  Hennessy-Milner logic \cite{HenMil85} (or equivalently, normal multi-modal logic)
  is here defined as classical propositional logic extended with 
  join-preserving diamonds.
  This is achieved by defining $\Lhm\colon \cat{BA} \to\cat{BA}$ as follows:
  For a Boolean algebra $B$, $\Lhm B$ is the free Boolean algebra generated
  by the set $\{ \diam{a}b \mid b \in B, a \in A \}$ modulo the congruence
  generated by the usual diamond equations, i.e.,
  $$
    \diam{a}\bot = \bot
    \quad\text{and}\quad
    \diam{a}(\phi_1 \lor \phi_2) = \diam{a}\phi_1 \lor \diam{a}\phi_2
  $$
  for all $a \in A$.
  The modal semantics $\rhohm \colon \Lhm\Pba \to \Pba(\Pow(-)^A)$
  is essentially the Boolean extension of $\rhotr$. In particular,
  $\rhohm_X(\diam{a}U) = \{ t \in \Pow(X)^A \mid t(a) \cap U \neq \emptyset\}$.
\end{example}

  The above description of Hennessy-Milner logic is a special case of a
  more general approach described in the next example.

\begin{example}\label{exm:pl}
  If $\cat{A}$ in the dual adjunction is a variety of algebras,
  we can define a logic $(\fun{L}, \rho)$ for $\fun{T} : \cat{C} \to \cat{C}$ by
  \emph{predicate liftings and axioms} as in \cite[Definition~4.2]{KupKurPat04}
  and \cite[Theorems 4.7 and 8.8]{KurRos12}.
  An \emph{$n$-ary predicate lifting} is a natural transformation
  $$
    \lambda : \fun{UP}^n \to \fun{UPT},
  $$
  where $\fun{P}^nX$ is the $n$-fold product of $\fun{P}X$ in $\cat{A}$ and
  $\fun{U} : \cat{A} \to \cat{Set}$ is the forgetful functor.
  Together with a suitable notion of \emph{axioms},
  a collection $\Lambda$ of such predicate liftings yields a functor
  $
    \fun{L} : \cat{A} \to \cat{A}
  $
  sending $A \in \cat{A}$ to the free algebra generated by
  $\{ \und{\lambda}(a_1, \ldots, a_n) \mid \lambda \in \Lambda, a_i \in A \}$
  modulo (instantiations of) the axioms.
  Define $\rho : \fun{LP} \to \fun{PT}$ on generators by
  $\rho_X(\und{\lambda}(a_1, \ldots, a_n)) = \lambda_X(a_1, \ldots, a_n) \in \fun{PT}X$.
  If $\rho$ is well-defined then it is natural transformation and 
  $(\fun{L}, \rho)$ is a logic for $\cat{Coalg}(\fun{T})$.
  All logics in e.g. \cite{BakHan17,BezGroVen19-report,KupPat11} are instances hereof.
\end{example}

  Next, we interpret positive modal logic \cite{Dun95,CelJan99},
  whose coalgebraic semantics over posets can be found in \cite[Example 2.4]{KapKurVel14},
  in topological spaces:

\begin{example}\label{exm:pml-top}
  Consider the dual adjunction
  $
    \begin{tikzcd}[cramped, sep=1.5em]
      \fun{\Omega} : \cat{Top}
            \arrow[r, shift left=1.7pt]
        & \cat{DL} : \fun{pf}
            \arrow[l, shift left=1.7pt]
    \end{tikzcd}
  $,
  where $\fun{\Omega}$ takes open subsets of a topological space, viewed as a 
  distributive lattice, and $\fun{pf}$ takes prime filters of a distributive
  lattice topologised by the subbase $\{ \tilde{a} \mid a \in A \}$, 
  where $\tilde{a} = \{ p \in \fun{pf}A \mid a \in p \}$.
  The Vietoris functor $\fun{V} : \cat{Top} \to \cat{Top}$ takes
  $X \in \cat{Top}$ to its collection of compact subsets topologised by the subbase consisting of
  $\dbox a = \{ c \in \fun{V}X \mid c \subseteq a \}$ and
  $\ddiamond a = \{ c \in \fun{V}X \mid c \cap a \neq \emptyset \}$,
  where $a$ ranges over the opens of $X$.
  For a continuous map $f : X \to X'$ the map $\fun{V}f$ takes 
  direct images.

  Positive modal logic is given by the functor $\fun{N} : \cat{DL} \to \cat{DL}$
  that sends a distributive lattice $A$ to the free distributive lattice
  generated by the set $\{ \Box a, \Diamond a \mid a \in A \}$ modulo the axioms
  \begin{align*}
    \Box\top &= \top
      & \Diamond\bot &= \bot \\
    \Box a \wedge \Box b &= \Box(a \wedge b)
      & \Diamond a \vee \Diamond b &= \Diamond(a \vee b) \\
    \Diamond a \wedge \Box b &\leq \Diamond(a \wedge b)
      & \Box(a \vee b) &\leq \Box a \vee \Diamond b
  \end{align*}
  The interpretation of this logic in $\fun{V}$-coalgebras is given by
  the natural transformation $\rho : \fun{N\Omega} \to \fun{\Omega V}$,
  defined on generators by $\Box a \mapsto \dbox a$ and $\Diamond a \mapsto \ddiamond a$.
\end{example}

  We now recall \emph{linear weighted automata}, see e.g.~\cite[Section 3.2]{BonEA12}.
  This is particularly interesting because it is an example of a 
  \emph{many-valued} logic, with truth values in some field $\Bbbk$.

\begin{example}\label{exm:trace-vec}
  Let $\Bbbk$ be a field and
  let $\cat{Vec}_{\Bbbk}$ be the category of vector spaces over $\Bbbk$.
  For a set $A$ of labels, define the endofunctor $\fun{W}$ on $\cat{Vec}_{\Bbbk}$
  by $\fun{W} = \Bbbk \times (-)^A$, where $(-)^A$ is the collection of
  maps $A \to (-)$ with a pointwise vector space structure.
  Then \emph{linear weighted automata} are $\fun{W}$-coalgebras.
  
  We wish to interpret \emph{linear trace logic} in such coalgebras,
  that is, formulae in the language given by the grammar
  $$
    \phi ::= p \mid \langle a \rangle \phi
  $$
  In order to do this in the abstract coalgebraic framework,
  we use the dual adjunction between $\cat{Vec}_{\Bbbk}$ and $\cat{Set}$.
  In one direction this is given by the hom-functor
  $(-)^{\circ} = \Hom(-, \Bbbk) : \cat{Vec}_{\Bbbk} \to \cat{Set}$.
  Conversely, for a set $X$ define $X^{\wedge}$ to be the collection
  $\Hom(X, \Bbbk)$ with pointwise vector space structure.
  It is easy to see that this yields a functor $\cat{Set} \to \cat{Vec}_{\Bbbk}$.

  The interpretation of $p$ is given by the nullary predicate lifting
  $\lambda^p \in (\fun{W}-)^{\circ}$ given by
  $\lambda^p_X : \fun{W}X \to \Bbbk : (r, t) \mapsto r$.
  Then $\llb p \rrb = \lambda^p_X \circ \gamma : X \to \Bbbk$.
  The interpretation of the diamonds is given by the unary predicate lifting
  $\lambda^{\langle a \rangle} : \fun{U}(-)^{\circ} \to \fun{U}(\fun{W}-)^{\circ}$
  defined by
  $$
    \lambda^{\langle a \rangle}_X(m) : \fun{W}X \to \Bbbk : (r, t) \mapsto m(t(a)).
  $$
  (Note that in this case $\fun{U}$ is the identity functor on $\cat{Set}$.)
  Concretely, this means that if $\llb p \rrb_{\gamma}(y) = r \in \Bbbk$
  and there is an $a$-transition $x \overset{a}{\too} y$, then
  $\llb \langle a \rangle p \rrb(x) = r$.
  
\end{example}
  
  Since for $V \in \cat{Vec}_{\Bbbk}$ the set $\Hom(V, \Bbbk)$
  forms a vector space, rather than just a set,
  we can also interpret vector space operations in $\fun{W}$-coalgebras.
  We make this modification in the following example.

\begin{example}  \label{exm:mod-vec}
  Let $\Bbbk$ be a field and
  $
  \begin{tikzcd}[cramped, sep=1.5em]
    \cat{Vec}_{\Bbbk}
          \arrow[r, shift left=1.7pt]
      & \cat{Vec}_{\Bbbk}
          \arrow[l, shift left=1.7pt]
  \end{tikzcd}
  $
  the dual adjunction between vector spaces over $\Bbbk$
  given in both directions by taking dual vector space via the contravariant functor
  $(-)^{\vee} = \Hom(-, \Bbbk) : \cat{Vec}_{\Bbbk} \to \cat{Vec}_{\Bbbk}$.
  (Note that the functors $(-)^{\circ}$ and $(-)^{\vee}$ are related via
  $(-)^{\circ} = \fun{U}_{\cat{Vec}} \circ (-)^{\vee}$,
  where $\fun{U}_{\cat{Vec}} : \cat{Vec} \to \cat{Set}$ is the forgetful functor.)
%
%
  We extend linear trace logic with vector space operations,
  and work with the language given by the grammar
  $$
    \phi ::= 0 \mid p \mid r \cdot \phi \mid \phi + \phi \mid \langle a \rangle\phi,
  $$
  where $a \in A$, $r \in \Bbbk$, and $p$ is a single proposition letter
  (the termination predicate).
  We refer to this \emph{linear Hennessy-Milner logic}.
  The interpretation of a formula $\phi$ in this (many-valued) setting
  is a linear map $\llb \phi \rrb : X \to \Bbbk$.
  The connectives $0$, $+$ and $r$ are interpreted via the corresponding operations
  in vector spaces, and for $p$ and $\langle a \rangle$ we use
  the predicate liftings from Example \ref{exm:trace-vec}.
%
%
  Together with the axioms
  $\langle a \rangle(\phi + \psi) = \langle a \rangle \phi + \langle a \rangle \psi$
  and $r \cdot \langle a \rangle \phi = \langle a \rangle (r \cdot \phi)$
  this gives rise to an endofunctor $\fun{L} : \cat{Vec}_{\Bbbk} \to \cat{Vec}_{\Bbbk}$,
  and a logic $(\fun{L}, \rho)$ for linear weighted automata.
  One can show that logical equivalence coincides with language semantics
  if the state-space is finite-dimensional.
\end{example}

\paragraph{Relations as jointly mono spans}
  We are interested in giving certain relations a special status.
  In $\cat{Set}$, a binary relation $B \subseteq X \times X$ corresponds
  to an injective map $B \injr X \times X$.
  This generalises to an arbitrary category (possibly lacking products)
  via the notion of a \emph{jointly mono span}: 
  A span 
  $
  \begin{tikzcd}[cramped, sep=1.5em]
    X_1
      & B \arrow[l, swap, "\pi_1"] \arrow[r, "\pi_2"] 
      & X_2
  \end{tikzcd}
  $
  in a category $\cat{C}$ is called \emph{jointly mono} if for all
  $\cat{C}$-arrows $h, h'$ with codomain $B$ it satisfies:
  $
    \text{if } \pi_1\circ h = \pi_1\circ h' \text{ and } \pi_2\circ h = \pi_2\circ h' \text{ then } h = h'.
  $
  We sometimes write the above span as $(B,\pi_1,\pi_2)$, leaving codomains implicit.
  If $\cat{C}$ has products, then $(B, \pi_1, \pi_2)$ is a jointly mono span if and 
  only if the pairing $\langle \pi_1, \pi_2 \rangle : B \to X_1 \times X_2$ is 
  monic.

  The collection of jointly mono spans between two objects $X_1, X_2 \in \cat{C}$
  can be ordered as follows: $(B, \pi_1, \pi_2) \leq (B', \pi_1', \pi_2')$
  if there exists a (necessarily monic) map $k : B \to B'$ such that $\pi_i = \pi_i' \circ k$.
  If
  $(B, \pi_1, \pi_2) \leq (B', \pi_1', \pi_2')$ \emph{and}
  $(B', \pi_1', \pi_2') \leq (B, \pi_1, \pi_2)$, then the two spans must be
  isomorphic. We write $\cat{Rel}(X_1, X_2)$ for the poset of jointly mono spans
  between $X_1$ and $X_2$ up to isomorphism.

\paragraph{Image factorisations and regular epis}
  We will also need a generalisation of image factorisation.
  A category $\cat{C}$ is said to have \emph{$(\ms{E}, \ms{M})$-factorisations}
  for some classes $\ms{E}$ and $\ms{M}$ of $\cat{C}$-morphisms, 
  if every morphism $f \in \cat{C}$ factorises as $f = m \circ e$ with
  $e \in \ms{E}$ and $m \in \ms{M}$. We say that $\cat{C}$ has an
  \emph{$(\ms{E}, \ms{M})$-factorisation system}
  \cite[Definition~14.1]{AdaHerStr90} if moreover both $\ms{E}$ and $\ms{M}$
  are closed under composition, and whenever
  $g \circ e = m \circ f$, with $e \in \ms{E}$ and $m \in \ms{M}$,
  there exists a unique diagonal fill-in $d$ such that $f = d\circ e$ and $g = m\circ d$.
  In a diagram:
  $$
    \begin{tikzcd}
      {}    \arrow[r, two heads, "e"]
            \arrow[d, "f" swap]
        & [.5em] {}
            \arrow[d, "g"]
            \arrow[dl, dashed, "d", shorten <=4pt, shorten >=4pt] \\ [.5em]
      {}    \arrow[r, >->, "m" swap]
        & {}
    \end{tikzcd}
  $$
 
  An epi $e$ is \emph{regular} if it is a coequalizer.
  In a variety, the regular epis are precisely the surjective morphisms.
  The categories $\cat{Set}, \cat{Pos}, \cat{Top}, \cat{Vec}, \cat{SL}, \cat{Stone}$
  all have a $(\ms{R}eg\ms{E}pi, \ms{M}ono)$-factorisation system.
  

\section{Logic-induced bisimulations}
\label{sec:rho-bis}

  We are now ready to define our logic-induced notion of bisimulation.
  Throughout this section, we fix a dual adjunction
  $
  \begin{tikzcd}[sep=1.5em, cramped]
    \fun{P} : \cat{C}
          \arrow[r, shift left=1.7pt]
      & \cat{A} : \fun{S}
          \arrow[l, shift left=1.7pt]
  \end{tikzcd}
  $,
  an endofunctor $\fun{T}$ on $\cat{C}$, and a logic $(\fun{L}, \rho)$
  for $T$-coalgebras.
  Moreover, we assume that $\cat{C}$ has pull\-backs and, in addition,
  that $\cat{A}$ has pullbacks or $\cat{C}$ has pushouts.
  Both conditions hold in all examples given in Section \ref{sec:cml-rel}.
  In particular, if $\cat{A}$ is variety of algebras then pullbacks exist and
  are computed as in $\cat{Set}$.

\subsection{Definition and first examples}\label{subsec:def}

  The basic ingredient for the definition of $\rho$-bisimulation is
  the notion of a \emph{dual span}: A jointly mono span
  $
  \begin{tikzcd}[cramped, sep=1.5em]
    X_1 
      & B \arrow[l, swap, "\pi_1"] \arrow[r, "\pi_2"]
      & X_2
  \end{tikzcd}
  $
  in $\cat{C}$
  is mapped by $\fun{P}$ to a cospan
  $
  \begin{tikzcd}[cramped, sep=1.5em]
    \fun{P}X_1
      & \fun{P}B
          \arrow[l, latex-, swap, "\fun{P}\pi_1"] 
          \arrow[r, latex-, "\fun{P}\pi_2"] 
      & \fun{P}X_2
  \end{tikzcd}
  $
  in $\cat{A}$.
  Taking its pullback we obtain a jointly mono span in $\cat{A}$,
  which we denote by  $(\dpo{B}, \dpo{\pi}_1, \dpo{\pi}_2)$ and refer to as
  the \emph{dual span} of $(B, \pi_1, \pi_2)$. In a diagram:
  \begin{equation*}
    \begin{tikzcd}[row sep=0em, column sep=1.5em]
        & \dpo{B}
            \arrow[dl, bend right=5, "\ov{\pi}_1" {swap,pos=.4}]
            \arrow[dr, bend left=5, "\ov{\pi}_2" {pos=.4}]
        & \\
      \fun{P}X_1
            \arrow[dr, bend right=5, "\fun{P}\pi_1" swap] 
        & 
        & \fun{P}X_2
            \arrow[dl, bend left=5, "\fun{P}\pi_2"] \\ [-.3em]
        & \fun{P}B 
        &
    \end{tikzcd}
  \end{equation*}
  If $\cat{C}$ has pushouts, dual spans exist because dual adjoints send
  pushouts to pullbacks.
  In the classic case where $\fun{P} = \Pba\colon \cat{Set} \to\cat{BA}$
  maps a set to its Boolean algebra of subsets,
  the dual span $(\dpo{B}, \ov{\pi}_1, \ov{\pi}_2)$ consists of
  \emph{$B$-coherent pairs (of subsets of $X_1$ and $X_2$)},
  that is, pairs $(a_1, a_2) \in \fun{P}X_1 \times \fun{P}X_2$ of subsets
  satisfying $B[a_1] \subseteq a_2$ and $B^{-1}[a_2] \subseteq a_1$.
  This notion of $B$-coherent pairs has been used in the definitions of
  $\Lambda$-bisimulation~\cite{BakHan17},
  neighbourhood bisimulation \cite{HanKupPac09},
  and conditional bisimulation \cite{BalCin18}.
  
  We proceed to the definition of a $\rho$-bisimulation.

\begin{definition}\label{def:bisim}
  Let $\gamma_1 \colon X_1 \to \fun{T}X_1$ and $\gamma_2 \colon X_2 \to \fun{T}X_2$ be $\fun{T}$-coalgebras.
  A jointly mono span
  $
  \begin{tikzcd}[cramped, sep=1.5em]
    X_1 
      & B \arrow[l, swap, "\pi_1"] \arrow[r, "\pi_2"]
      & X_2
  \end{tikzcd}
  $
  is a \emph{$\rho$-bisimulation between $\gamma_1$ and $\gamma_2$} if
  \begin{equation}\label{eq:rho-bis}
    \fun{P}\pi_1 \circ \gamma_1^* \circ \fun{L}\ov{\pi}_1 
      = \fun{P}\pi_2 \circ \gamma_2^* \circ \fun{L}\ov{\pi}_2.
  \end{equation}
\end{definition}      

Definition~\ref{def:bisim} 
is structural in the sense that it is defined in terms of the coalgebra structure and the one-step modal semantics  $\rho$ (via the complex algebras $\gamma_i^*$). In particular, it does not refer to the collection of all formulae nor to the initial $\fun{L}$-algebra.
Equation~\eqref{eq:rho-bis} provides a coherence condition that can be checked in concrete settings. We provide examples below.
First, we give a more conceptual characterisation in terms of dual spans.

\begin{proposition}\label{prop:pb-alg   }
  A jointly mono span $(B,\pi_1,\pi_2)$         
  is a $\rho$-bisimulation between $(X_1, \gamma_1)$ and $(X_2, \gamma_2)$
  if and only if its dual span $(\dpo{B},\ov{\pi}_1,\ov{\pi}_2)$ 
  is a congruence between $\gamma_1^*$ and $\gamma_2^*$.
\end{proposition}
\begin{proof}
  Suppose $(B, \pi_1, \pi_2)$ is a $\rho$-bisimulation,
  i.e., equation \eqref{eq:rho-bis} holds true. Then the
  outer shell of the diagram in \eqref{eq:def-bisim} commutes and
  the universal property of the pullback $\dpo{B}$ yields a morphism
  $\beta : \fun{L}\dpo{B} \to \dpo{B}$ such that all squares in
  \eqref{eq:def-bisim} commute.
  \begin{equation}\label{eq:def-bisim}
    \begin{tikzcd}[row sep=0em, column sep=1.5em]
        & \fun{L}\dpo{B}
            \arrow[dl, bend right=5, "\fun{L}\ov{\pi}_1" swap]
            \arrow[dr, bend left=5, "\fun{L}\ov{\pi}_2"]
            \arrow[dd, dashed, "\beta"]
        & \\
      \fun{LP}X_1
            \arrow[dd, "\gamma_1^*" left] 
        &
        & \fun{LP}X_2
            \arrow[dd, "\gamma_2^*" right] \\
        & \dpo{B}
            \arrow[dl, bend right=5, "\ov{\pi}_1" {swap,pos=.4}]
            \arrow[dr, bend left=5, "\ov{\pi}_2" {pos=.4}]
        & \\
      \fun{P}X_1
            \arrow[dr, bend right=5, "\fun{P}\pi_1" swap] 
        & 
        & \fun{P}X_2
            \arrow[dl, bend left=5, "\fun{P}\pi_2"] \\ [-.3em]
        & \fun{P}B 
        &
    \end{tikzcd}
  \end{equation}
  Conversely, the existence of such a $\beta$ making the inner squares
  commute implies commutativity of the outer shell of the diagram.
\end{proof}
    
We instantiate the definition 
for some of the examples of Section~\ref{sec:cml-rel}.

\begin{example}\label{exm:rho-bis-lambda}
  Recall the setting of Example~\ref{exm:pl} where $\cat{A}$ is a variety and
  $(\fun{L}, \rho)$ is given by predicate liftings and axioms,
  and let $(X_1, \gamma_1)$ and $(X_2, \gamma_2)$ be two $\fun{T}$-coalgebras.
  If $\cat{C}$ is concrete, then a jointly mono span $\Bspan$
  is a $\rho$-bisimulation between $(X_1,\gamma_1)$ and $(X_2,\gamma_2)$
  if for all $(x_1, x_2) \in B$, $\lambda \in \Lambda$
  and all $B$-coherent $(a_1, a_2) \in \fun{P}X_1 \times \fun{P}X_2$ we have:
  \[
    \gamma_1(x_1) \in \lambda_{X_1}(a_1)
    \quad\text{iff}\quad
    \gamma_2(x_2) \in \lambda_{X_2}(a_2).
  \]
  The notion of a $\rho$-bisimulation thus generalises that of a
  $\Lambda$-bisimulation from \cite{BakHan17,BezGroVen19-report},
  where $\Lambda$ denotes a collection of (open) predicate liftings.            
  Examples~\ref{exm:pml-top-bis}, \ref{exm:mod-vec-bis} and \ref{exm:bis-join-logic}
  below are instances hereof.
\end{example}

\begin{example}\label{exm:pml-top-bis}
  In the setting of positive modal logic from Example~\ref{exm:pml-top},
  a $\rho$-bisimulation between $(X_1, \gamma_1)$
  and $(X_2, \gamma_2)$ is a subspace $B \subseteq X_1 \times X_2$
  with projections $\pi_i : B \to X_i$ satisfying for all $(x_1, x_2) \in B$
  and all $B$-coherent pairs of opens $(a_1, a_2) \in \fun{\Omega}X_1 \times \fun{\Omega}X_2$:
  \begin{center}
    $\gamma_1(x_1) \subseteq a_1$ iff $\gamma_2(x_2) \subseteq a_2$
    \quad and \quad
    $\gamma_1(x_1) \cap a_1 \neq \emptyset$ iff $\gamma_2(x_2) \cap a_2 \neq \emptyset$.
  \end{center}
\end{example}

\begin{example}\label{exm:mod-vec-bis}
  In the setting of linear Hennessy-Milner logic from Example~\ref{exm:mod-vec},
  a jointly mono span between the state-spaces of $\fun{W}$-coalgebras
  $(X_1, \gamma_1)$ and $(X_2, \gamma_2)$ is a linear subspace of $X_1 \times X_2$. 
  The dual span of $\Bspan$ is the linear subspace of $X_1^{\vee} \times X_2^{\vee}$
  consisting of those pairs of $\Bbbk$-valued, linear predicates
  $(h_1, h_2) \in X_1^{\vee} \times X_2^{\vee}$
  such that $(x_1, x_2) \in B$ implies $h_1(x_1) = h_2(x_2)$.
  Unravelling the definitions shows that $(B, \pi_1, \pi_2)$ is 
  a $\rho$-bisimulation between $(X_1, \gamma_1)$ and $(X_2, \gamma_2)$,
  if for all $(x_1, x_2) \in B$, we have
  $\llb p \rrb_{\gamma}(x_1) = \llb p \rrb_{\gamma}(x_2)$, and:
  \begin{center}
    if $x_1 \overset{a}{\too} y_1$ and $x_2 \overset{a}{\too} y_2$,
          then $h_1(y_1) = h_2(y_2)$ for all $(h_1, h_2) \in \dpo{B}$.
  \end{center}
\end{example}

\begin{example}\label{exm:id-rho-bis}
  More abstractly, suppose given any logical connection
  $
  \begin{tikzcd}[sep=1.5em, cramped]
    \fun{P} : \cat{C}
          \arrow[r, shift left=1.7pt]
      & \cat{A} : \fun{S}
          \arrow[l, shift left=1.7pt]
  \end{tikzcd}
  $,
  functor $\fun{T} : \cat{C} \to \cat{C}$ and logic $(\fun{L}, \rho)$ for
  $\fun{T}$-coalgebras.
  Then for every $\fun{T}$-coalgebra $(X, \gamma)$ the
  jointly mono span $(X, \id_X, \id_X)$ is a $\rho$-bisimulation on $(X, \gamma)$.
\end{example}

  We complete this subsection by showing that the notion of a $\rho$-bisimulation
  is adequate, that is, $\rho$-bisimulations preserve truth.
  We say that a span $(B, \pi_1, \pi_2)$ between $(X_1, \gamma_1)$ and
  $(X_2, \gamma_2)$ is \emph{truth preserving}
  if $\th_{\gamma_1} \circ\; \pi_1 = \th_{\gamma_2} \circ\; \pi_2$.
  If $\cat{C}$ is concrete, this means that if $(x_1,x_2) \in B$
  then $x_1$ and $x_2$ have the same theory, i.e., satisfy the same formulae.

\begin{proposition}\label{prop:adequacy}
  If
   $
  \begin{tikzcd}[cramped, sep=1.5em]
    X_1
      & B \arrow[l, swap, "\pi_1"] \arrow[r, "\pi_2"]
      & X_2
  \end{tikzcd}
  $
  is a $\rho$-bisimulation between $\fun{T}$-coalgebras
  $(X_1, \gamma_1)$ and $(X_2, \gamma_2)$,
  then $\th_{\gamma_1} \circ\, \pi_1 = \th_{\gamma_2} \circ\, \pi_2$.
\end{proposition}
\begin{proof}
  Let $\beta : \fun{L}\dpo{B} \to \dpo{B}$ be given as in \eqref{eq:def-bisim},
  and let $h_\beta\colon \Phi \to \dpo{B}$ be the unique morphism from the initial
  $\fun{L}$-algebra to $(\dpo{B}, \beta)$.
  By construction of $\beta$,
  $\dpo{\pi}_i\colon (\dpo{B},\beta) \to (\fun{P}X_i,\gamma_i^*)$ are $\fun{L}$-algebra morphisms.
  By uniqueness of initial morphisms,
  $\sem{-}_{\gamma_i} = \dpo{\pi}_i \circ h_{\beta}$,
  and hence
  $\fun{S}\sem{-}_{\gamma_i} = \fun{S}h_\beta \circ \fun{S}\dpo{\pi}_i$, for $i = 1, 2$.
  Combining this with $\fun{S}\dpo{\pi}_1 \circ \fun{SP}\pi_1 = \fun{S}\dpo{\pi}_2 \circ \fun{SP}\pi_2$ (obtained by applying $\fun{S}$ to the pullback square of $(\dpo{B},\dpo{\pi}_1,\dpo{\pi}_2)$),
  it follows that
  $\fun{S}\llb \cdot \rrb_{\gamma_1} \circ \fun{SP}\pi_1
  = \fun{S}\llb \cdot \rrb_{\gamma_2} \circ \fun{SP}\pi_2$.
  Recall that the theory map is the adjoint of the semantic map, i.e.,
  $\th_{\gamma_i} = \fun{S}\sem{-}_{\gamma_i} \circ \unitc_{X_i}$
  where $\unitc:\Id_{\cat{C}}\to\fun{SP}$ is a unit of the logical connection
  $
  \begin{tikzcd}[sep=1.5em, cramped]
    \fun{P} : \cat{C}
          \arrow[r, shift left=1.7pt]
      & \cat{A} : \fun{S}
          \arrow[l, shift left=1.7pt]
  \end{tikzcd}
  $.
  It then follows from naturality of $\unitc$ that:
  \begin{alignat*}{3}
    \th_{\gamma_1} \circ\; \pi_1
      &= \fun{S}\llb \cdot \rrb_{\gamma_1} \circ \unitc_{X_1} \circ \pi_1
      &&= \fun{S}\llb \cdot \rrb_{\gamma_1} \circ \fun{SP}\pi_1 \circ \unitc_B
      && \\
      &= \fun{S}\llb \cdot \rrb_{\gamma_2} \circ \fun{SP}\pi_2 \circ \unitc_B
      &&= \fun{S}\llb \cdot \rrb_{\gamma_2} \circ \unitc_{X_2} \circ \pi_2
      &&= \th_{\gamma_2} \circ\; \pi_2
  \end{alignat*}
  as desired.
\end{proof}

\subsection{Lattice structure and composition of $\rho$-bisimulations}\label{subsec:lat}

  In the remainder of Section~\ref{sec:rho-bis}
  we assume that $\cat{C}$ is finitely complete and well-powered,
  hence $\cat{Rel}(X_1, X_2)$ is simply the poset of subobjects of $X_1\times X_2$.
  Besides, assume that $\cat{C}$ has an $(\ms{E}, \ms{M})$-factorisation system
  with $\ms{M} = \ms{M}ono$.
  Again, all examples in Section \ref{sec:cml-rel} satisfy these assumptions.
  
  It is well known that bisimulations for $\cat{Set}$-based coalgebras
  are closed under composition if and only if 
  the coalgebra functor preserves weak pullbacks \cite{Rut00}.
  We know from \cite[Example~3.3]{BakHan17} that $\Lambda$-bisimulations
  do not always compose, even for weak pullback-preserving functors,
  so as a consequence of Example~\ref{exm:rho-bis-lambda} the same failure
  occurs for $\rho$-bisimulations.
  However, in special cases we \emph{can} compose.
  Let us first define what we mean by the composition of two relations.

\begin{definition}\label{def:comp-rel}
  The composition of two jointly mono spans $(B, \pi_1, \pi_2)$ in $\cat{Rel}(X_1, X_2)$
  and $(B', \pi_2', \pi_3)$ in $\cat{Rel}(X_2, X_3)$ is given as follows:
  The pullback $(C, c_1, c_3)$ of $\pi_2$ and $\pi_2'$ yields projections
  $\pi_i\circ c_i : C \to X_i$, and we
  define $B \circ B'$ via the $(\ms{E}, \ms{M}ono)$-factorisation of
  $\langle \pi_1\circ c_1, \pi_3\circ c_3 \rangle$:
  $$
  \begin{tikzcd}[]
    C     \arrow[dr, ->>, bend right=5]
          \arrow[rr, "{\langle \pi_1\circ c_1, \pi_3\circ c_3 \rangle}"]
      & [1em]
      & [1em]
        X_1 \times X_3. \\ [-2em]
      & B \circ B'
          \arrow[ur, hook, bend right=5]
      &
  \end{tikzcd}
  $$
\end{definition}

  Call a $\rho$-bisimulation \emph{full} if both projections are split epi, that is, they
  have a section.
  For $\cat{Set}$-based coalgebras this means that 
  the projections are surjective, i.e.,
  each state in $(X_1,\gamma_1)$ is $\rho$-bisimilar to some state in $(X_2,\gamma_2)$,
  and vice versa.

\begin{lemma}\label{lem:rho-bis-span}
  Let
  $
  \begin{tikzcd}[cramped, sep=1.5em]
    X_1
      & S \arrow[l, swap, "\zeta_1"] \arrow[r, "\zeta_2"]
      & X_2
  \end{tikzcd}
  $
  and
  $
  \begin{tikzcd}[cramped, sep=1.5em]
    X_1
      & B \arrow[l, swap, "\pi_1"] \arrow[r, "\pi_2"]
      & X_2
  \end{tikzcd}
  $
  be spans between $\fun{T}$-coalgebra $(X_1, \gamma_1)$ and $(X_2, \gamma_2)$
  and suppose $e : S \to B$
  is an epimorphism such that $\zeta_i = \pi_i \circ e$.
  Then $(S, \zeta_1, \zeta_2)$ satisfies \eqref{eq:rho-bis} if and only if
  $(B, \pi_1, \pi_2)$ does.
\end{lemma}
\begin{proof}
  Since $S \to B$ is epic $\fun{P}B \to \fun{P}S$ is monic.
  It follows that the pullback of $(S, \zeta_1, \zeta_2)$ coincides with
  the pullback to $(B, \pi_1, \pi_2)$. With this observation the
  proof of the proposition follows from a straightforward computation.
\end{proof}

  Now we can show that full bisimulations compose.

\begin{proposition}\label{prop:comp}
  Full bisimulations are closed under composition.
\end{proposition}
\begin{proof}
  Let $(B, \pi_1, \pi_2)$ be a $\rho$-bisimulation between $(X_1, \gamma_1)$ and
  $(X_2, \gamma_2)$, and $(B', \pi_2', \pi_3)$ a $\rho$-bisimulation between
  $(X_2, \gamma_2)$ and $(X_3, \gamma_3)$.
  Write $(C, c_1, c_3)$ for the pullback of $\pi_2$ and $\pi_2'$
  and define $(\dpo{C}, \dpo{c}_1, \dpo{c}_3)$ to be the pullback
  of $\dpo{\pi}_2$ and $\dpo{\pi}_2'$.
  By Lemma \ref{lem:rho-bis-span} it suffices to show that
  $(C, \pi_1c_1, \pi_3c_3)$ satisfies the $\rho$-bisimulation condition.

  Since all the $\pi_i$ are split epic, so are $c_1$ and $c_3$
  (cf.~Lemma \ref{lem-pb-split}).
  According to Lemma \ref{lem-pb-po}
  this implies that the pullback square
  \begin{equation*}
    \begin{tikzcd}[row sep=.2em]
        & C
          \arrow[dl, "c_1" swap]
          \arrow[dr, "c_2"]
        & \\
      B_1 \arrow[dr, "\pi_2" swap]
        &
        & B_2 \arrow[dl, "\pi_2'"] \\
        & X_2
        &
    \end{tikzcd}
  \end{equation*}
  is also a pushout. 
  Therefore square \circled{4} below is a pullback, while \circled{1}, \circled{2}
  and \circled{3} are pullbacks by definition.
  It follows from repeated application of the pullback lemma that the
  outer square is a pullback as well.
  \begin{equation}\label{eq:pbs}
  \begin{tikzcd}[row sep=.2em]
      &
      & \dpo{C}
          \arrow[dl, "\ov{c}_1" swap] \arrow[dr, "\ov{c}_3"]
          \arrow[dd, phantom, "\circled{1}" pos=.5]
      &
      & \\
      & \dpo{B}_1
          \arrow[dl, "\ov{\pi}_1" swap]
          \arrow[dr, "\ov{\pi}_2"]
          \arrow[dd, phantom, "\circled{2}"]
      &
      & \dpo{B}_2
          \arrow[dl, "\ov{\pi}_2'" swap]
          \arrow[dr, "\ov{\pi}_3"]
          \arrow[dd, phantom, "\circled{3}"]
      & \\
    \fun{P}X_1
          \arrow[dr, "\fun{P}\pi_1" swap]
      &
      & \fun{P}X_2
          \arrow[dl, "\fun{P}\pi_2" swap]
          \arrow[dr, "\fun{P}\pi_2'"]
          \arrow[dd, phantom, "\circled{4}"]
      &
      & \fun{P}X_3 \arrow[dl, "\fun{P}\pi_3"] \\
      & \fun{P}B_1 \arrow[dr, "\fun{P}c_1" swap]
      &
      & \fun{P}B_2 \arrow[dl, "\fun{P}c_3"]
      & \\
      &
      & \fun{P}C
      &
      &
  \end{tikzcd}
  \end{equation}
  As a consequence
  $(\ov{\pi}_1\ov{c}_1, \ov{\pi}_3\ov{c}_3)$ is jointly monic.
  
  In order to show that $(C, \pi_1c_1, \pi_3c_3)$ is a $\rho$-bisimulation
  we need to prove
  \begin{equation}\label{eq:composition}
    \fun{P}c_1 \circ \fun{P}\pi_1 \circ \gamma_1^*
        \circ \fun{L}\ov{\pi}_1 \circ \fun{L}\ov{c}_1
      = \fun{P}c_3 \circ \fun{P}\pi_3 \circ \gamma_3^*
        \circ \fun{L}\ov{\pi}_3 \circ \fun{L}\ov{c}_3
  \end{equation}
  This is the outer shell of the following diagram:
  $$
  
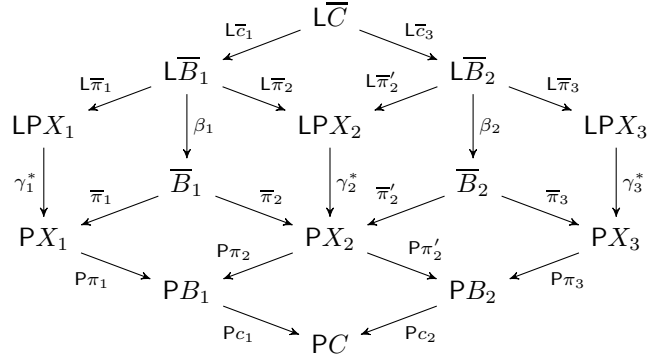
\begin{tikzcd}[row sep=tiny]
      &
      & \fun{L}\dpo{C}
          \arrow[dl, "\fun{L}\ov{c}_1" swap]
          \arrow[dr, "\fun{L}\ov{c}_3"]
      &
      & \\
      & \fun{L}\dpo{B}_1
          \arrow[dl, "\fun{L}\ov{\pi}_1" swap]
          \arrow[dr, "\fun{L}\ov{\pi}_2"]
          \arrow[dd, "\beta_1"]
      &
      & \fun{L}\dpo{B}_2
          \arrow[dl, "\fun{L}\ov{\pi}_2'" swap]
          \arrow[dr, "\fun{L}\ov{\pi}_3"]
          \arrow[dd, "\beta_2"]
      & \\
    \fun{LP}X_1 \arrow[dd, "\gamma_1^*" swap]
      &
      & \fun{LP}X_2 \arrow[dd, "\gamma_2^*"]
      &
      & \fun{LP}X_3 \arrow[dd, "\gamma_3^*"] \\
      & \dpo{B}_1 \arrow[dl, "\ov{\pi}_1" swap] \arrow[dr, "\ov{\pi}_2"]
      &
      & \dpo{B}_2 \arrow[dl, "\ov{\pi}_2'" swap] \arrow[dr, "\ov{\pi}_3"]
      & \\
    \fun{P}X_1 \arrow[dr, "\fun{P}\pi_1" swap]
      &
      & \fun{P}X_2 \arrow[dl, "\fun{P}\pi_2" swap] \arrow[dr, "\fun{P}\pi_2'"]
      &
      & \fun{P}X_3 \arrow[dl, "\fun{P}\pi_3"] \\
      & \fun{P}B_1 \arrow[dr, "\fun{P}c_1" swap]
      &
      & \fun{P}B_2 \arrow[dl, "\fun{P}c_2"]
      & \\
      &
      & \fun{P}C
      &
      &
  \end{tikzcd}
  $$
  Commutativity of the top square follows from applying $\fun{L}$ to cell
  \circled{1} in \eqref{eq:pbs}.
  The bottom square commutes by definition of $(C, c_1, c_3)$.
  All other squares commute because $(B_1, \pi_1, \pi_2)$ and 
  $(B_2, \pi_2', \pi_3)$ are assumed to be $\rho$-bisimulations.
  Thus \eqref{eq:composition} holds, and this ultimately proves that
  full $\rho$-bisimulations are closed under composition.
\end{proof}

  Another well-known result for bisimulations on $\cat{Set}$-coalgebras
  is that they form a complete lattice \cite{Rut00}.
  We now show that, provided $\cat{C}$
  has all coproducts,
  this also holds for $\rho$-bisimulations.     
  Recall that the empty coproduct $\coprod\emptyset =: \initobj$ is an initial object, 
  i.e., for all $C \in \cat{C}$ there is a unique morphism
  $!_C \colon \initobj \to C$.


\begin{definition}\label{def:union}
  The \emph{join} of a family of relations $(B_i,\pi_{i,1},\pi_{i,2})$, $i \in I$,
  in $\cat{Rel}(X_1, X_2)$, is the jointly mono span
  $\bigcup_{i\in I} B_i$ that arises from the factorisation
  \[
    \begin{tikzcd}[column sep=3.5em]
      \coprod_i B_i \arrow[r, two heads]
            \arrow[rr, bend left=15, shift left=2pt, "{\coprod_i \langle \pi_{i,1}, \pi_{i,2} \rangle }"]
        & \bigcup_i B_i \arrow[r, hook] 
        & X_1 \times X_2
    \end{tikzcd}
  \]
  The \emph{bottom element} $(I, \iota_1, \iota_2)$ in $\cat{Rel}(X_1, X_2)$
  is defined by the factorisation of the initial morphism:
  $
    \begin{tikzcd}[cramped, column sep=1.5em]
      \initobj \arrow[r, two heads]
        & I \arrow[r, hook,
                   "{\langle \iota_1, \iota_2 \rangle}"] 
        &[2em] X_1 \times X_2
    \end{tikzcd}
  $.
\end{definition}

  Indeed, $\bigcup_i B_i$ is an upper bound in $\cat{Rel}(X_1, X_2)$.
  Suppose $(B_i, \pi_{i,1}, \pi_{i,2}) \leq (S, s_1, s_2)$ for all $i$,
  then there are $t_i : B_i \to S$  such that $\pi_{i,j} = s_j \circ t_i$.
  From the coproduct we get $t : \coprod_{i \in I} B_i \to S$ and this makes
  the outer shell of the diagram below commute.
  $$
  \begin{tikzcd}[row sep=0em, column sep=1.7em]
      & \bigcup_i B_i
          \arrow[dr, hook]
          \arrow[dd, dashed, "d"]
      & \\
    \coprod_i B_i
          \arrow[ru, ->>]
          \arrow[rd, "t"]
      &
      & X_1 \times X_2 \\
      & S \arrow[ru, hook]
      &
  \end{tikzcd}
  $$
  The factorisation system now gives a diagonal $d : \bigcup_{i \in I} B_i \to S$ 
  witnessing that $S$ is bigger than $\bigcup_{i \in I}B_i$ in $\cat{Rel}(X_1, X_2)$.
  
\begin{example}
  Let $X_1$ and $X_2$ be objects in $\cat{C}$ and $(B_i, \pi_{i,1}, \pi_{i,2})$
  relations in $\cat{Rel}(X_1, X_2)$, where $i$ ranges over some index set $I$.
  View these as subobjects of $X_1 \times X_2$.
  We describe the join of all $B_i$.
  \begin{enumerate}
    \item If $\cat{C} = \cat{Set}$, $\cat{C} = \cat{Pos}$ or $\cat{C} = \cat{Top}$
          the join is given by the union in $X_1 \times X_2$.
    \item If $\cat{C} = \cat{Stone}$ then the join of the $B_i$ is
          the closure of $\bigcup B_i$ viewed as a subspace of $X_1 \times X_2$.
    \item If $\cat{C} = \cat{Vec}$ then the join of a family of relations
          $B_i \subseteq X_1 \times X_2$ is the smallest subspace of $X_1 \times X_2$
          containing $\bigcup B_i$. That is, $\bigvee B_i$ contains all
          vectors $v \in X_1 \times X_2$ of the form
          $v = v_{i_1} + \cdots + v_{i_n}$, with $v_{i_j} \in B_{i_j}$.
  \end{enumerate}
\end{example}

\begin{proposition}\label{prop:rel-jsl}
  If $\cat{C}$ has an $(\ms{E}, \ms{M}ono)$-factorisation system,
  binary products and all coproducts, then 
  the poset $\cat{Rel}(X_1,X_2)$ is a complete join-semilattice with
  join $\bigcup$ and bottom element $(I, \iota_1, \iota_2)$.
\end{proposition}
\begin{proof}
  Commutativity and associativity of the join follow from the
  fact that coproducts are commutative and associative.
  For idempotency note that for every $\Bspan$ in $\cat{Rel}(X_1,X_2)$
  we have an $(\ms{E}, \ms{M}ono)$-factorisation
  $
    \begin{tikzcd}[cramped, sep=1.5em]
      B + B \arrow[r, two heads, "\nabla"]
        &[.5em] B \arrow[r, hook]
        & X_1 \times X_2,
    \end{tikzcd}
  $
  where $\nabla$ is the codiagonal, so $B \cup B = B$.

  Next, we show that  $(I, \iota_1, \iota_2)$ is the bottom element in $(\cat{Rel}(X_1,X_2), \cup)$.
  That is, for all $\Bspan$ in $\cat{Rel}(X_1,X_2)$,
  $B \cup I$ is isomorphic to $B$. By the definition of a coproduct, 
  $$
  \begin{tikzcd}[]
    B     \arrow[rr, bend left=15, shift left=2pt, "i"]
          \arrow[r, "\cong" below]
      & \initobj + B
          \arrow[->>, r, "!_I + \id_B" below]
      &[1em] I + B
  \end{tikzcd}
  $$
  commutes, where $i$ is the inclusion that arises from the coproduct.
  Since $\ms{E}$ is closed under composition, the map 
  $i : B \to I + B$ is in $\ms{E}$.
  By definition of the join, the following commutes:
  \[
    \begin{tikzcd}[column sep=3.5em]
      B     \arrow[r, bend right=0, two heads]
            \arrow[rrr, bend left=12, shift left=1pt, hook, "{\langle \pi_1, \pi_2 \rangle}"]
        & I + B
            \arrow[r, two heads, shift right=2pt]
            \arrow[rr, bend right=15, shift right=2pt, "{[ \langle \iota_1, \iota_2 \rangle, 
                           \langle \pi_1  , \pi_2   \rangle ]}" below]
        & I \cup B \arrow[r, hook] 
        & X_1 \times X_2
    \end{tikzcd}
  \]
  Since factorisation systems are unique up to isomorphism, we get an isomorphism
  $B \cong B \cup I$.
\end{proof}

  We define \emph{$\rho$-bisimilarity} as the join of all $\rho$-bisimulations
  in $\cat{Rel}(X_1, X_2)$.
  The following proposition tells us that $\rho$-bisimilarity
  is itself a $\rho$-bisimulation.
  Given two $\fun{T}$-coalgebras
  $(X_1, \gamma_1)$ and $(X_2, \gamma_2)$, we denote by
  $\rho\hyphen\cat{Bis}(\gamma_1, \gamma_2)$ the sub-poset of
  $\cat{Rel}(X_1, X_2)$ of $\rho$-bisimulations
  between $(X_1, \gamma_1)$ and $(X_2, \gamma_2)$.

\begin{proposition}\label{prop:bis-jsl}
Under the assumptions of Proposition~\ref{prop:rel-jsl},
$\rho$-$\cat{Bis}(\gamma_1, \gamma_2)$ 
is closed under joins and bottom element in $\cat{Rel}(X_1, X_2)$.
Consequently, $\rho$-$\cat{Bis}(\gamma_1, \gamma_2)$
is a complete join-semilattice, and hence also a complete lattice.
\end{proposition}
\begin{proof}
  We give the proof for the binary case. This is easily adapted to arbitrary joins.
  
  As a consequence of Lemma \ref{lem:rho-bis-span} it suffices to 
  prove that the bisimulation condition holds for the span
  $
    \begin{tikzcd}[column sep=3em]
      X_1
        & B + S
            \arrow[l, "{[\pi_1, \sigma_1]}" {swap,pos=.4}]
            \arrow[r, "{[\pi_2, \sigma_2]}" {pos=.4}]
        & X_2.
    \end{tikzcd}
  $
  We write $(\dpo{B + S}, \dpo{[\pi_1,\sigma_1]}, \dpo{[\pi_2, \sigma_2]})$
  for its dual span.
  
  Since $\fun{P}$ is part of a dual adjunction it turns colimits into limits,
  hence $\fun{P}(B + S) = \fun{P}B \times \fun{P}S$ and
  $\fun{P}([\pi_i, \sigma_i]) = \langle \fun{P}\pi_i, \fun{P}\sigma_i \rangle$ for $i = 1, 2$.
  Denote by $\theta_B$ and $\theta_S$ the projections of
  $\fun{P}B \times \fun{P}S$ to $\fun{P}B$ and $\fun{P}S$.
  By the property of a product, in order to show that
  $\fun{P}([\pi_1, \sigma_1]) \circ \gamma_1^* \circ \fun{L}\dpo{[\pi_1, \sigma_1]}
    = \fun{P}([\pi_2, \sigma_2]) \circ \gamma_2^* \circ \fun{L}\dpo{[\pi_2, \sigma_2]}$ 
  holds, it suffices to show that
  \begin{equation}\label{eq-rb-coprod1}
    \theta_j \circ \fun{P}([\pi_1, \sigma_1]) \circ \gamma_1^* \circ \fun{L}\dpo{[\pi_1, \sigma_1]}
      = \theta_j \circ \fun{P}([\pi_2, \sigma_2]) \circ \gamma_2^*\circ \fun{L}\dpo{[\pi_2, \sigma_2]}
    \end{equation}
  for $j = B, S$.
  Since $\fun{P}([\pi_i, \sigma_i]) = \langle \fun{P}\pi_i, \fun{P}\sigma_i \rangle$ this
  reduces to proving
  \begin{align}
    \fun{P}\pi_1 \circ \gamma_1^* \circ \fun{L}\dpo{[\pi_1, \sigma_1]}
      &= \fun{P}\pi_2 \circ \gamma_2^* \circ \fun{L}\dpo{[\pi_2, \sigma_2]}
        \label{eq-rb-coprod2} \\
    \fun{P}\sigma_1 \circ \gamma_1^* \circ \fun{L}\dpo{[\pi_1, \sigma_1]}
      &= \fun{P}\sigma_2 \circ \gamma_2^* \circ \fun{L}\dpo{[\pi_2, \sigma_2]}.
        \label{eq-rb-coprod3}
  \end{align}
  We focus on the first equation, the second being similar.
  
  Let $(\dpo{B}, \dpo{\pi}_1, \dpo{\pi}_2)$ be the pullback of $(\fun{P}\pi_1, \fun{P}\pi_2)$.
  Since
  $
    \fun{P}\pi_1 \circ q_1
      = \theta_1 \circ \fun{P}(\pi_1 + \sigma_1) \circ q_1
      = \theta_1 \circ \fun{P}(\pi_2 + \sigma_2) \circ q_2
      = \fun{P}\pi_2 \circ q_2
  $,
  the triple $(\dpo{B + S}, \dpo{[\pi_1, \sigma_1]}, \dpo{[\pi_2, \sigma_2]})$
  forms 
  \begin{equation}
    \begin{tikzcd}[column sep=3.5em, row sep=1.4em, cramped]
        & \dpo{B + S} \arrow[dl, bend right=20, "\dpo{[\pi_1, \sigma_1]}" swap]
            \arrow[dr, bend left=10, "\dpo{[\pi_2, \sigma_2]}"] 
            \arrow[d, dashed, "h"]
        & \\
      \fun{P}X_1
            \arrow[dr, bend right=10, "\fun{P}(\pi_1 + \sigma_1)"] 
            \arrow[ddr, bend right=20, "\fun{P}\pi_1" swap]
        & \dpo{B}
            \arrow[l, "\dpo{\pi}_1" swap]
            \arrow[r, "\dpo{\pi}_2"]
        & \fun{P}X_2
            \arrow[dl, bend left=10, "\fun{P}(\pi_2 + \sigma_2)" swap]
            \arrow[ddl, bend left=20, "\fun{P}\pi_2"] \\
        & \fun{P}B \times \fun{P}S
            \arrow[d, "\theta_1"]
        & \\
        & \fun{P}B
        &
    \end{tikzcd}
  \end{equation}
  a cone of the pullback diagram of $\dpo{B}$, and 
  hence we get a mediating map $h : \dpo{B + S} \to \dpo{B}$ making the diagram
  commute.
  The equality in \eqref{eq-rb-coprod2} now follows from applying $\fun{L}$ to 
  this diagram and using that $B$ is a $\rho$-bisimulation.

  To see that $I$ is a $\rho$-bisimulation, 
  it suffices to show that $\initobj$ with the unique maps to $X_1$ and $X_2$ satisfies
  \eqref{eq:rho-bis}. This follows immediately from the fact that
  $\fun{P}(\initobj)$ is final in $\cat{D}$.
\end{proof}

While $\rho$-$\cat{Bis}(\gamma_1, \gamma_2)$ is
a complete sub-semilattice of $\cat{Rel}(X_1, X_2)$,
it need not inherit the meets.
This resembles the situation for Kripke bisimulations,
which are generally not closed under intersections.

\begin{example}
The categories $\cat{Set}$, $\cat{Top}$ and $\cat{Vec}_{\Bbbk}$ from Examples
\ref{exm:pml-top-bis}, \ref{exm:rho-bis-lambda} and \ref{exm:mod-vec-bis}
are well-powered, complete and cocomplete, and as mentioned in Section~\ref{sec:cml-rel}
have a $(\ms{R}eg\ms{E}pi, \ms{M}ono)$-factorisation system.
Hence $\rho$-bisimulations for positive modal logic, 
linear Hennessy-Milner logic and
coalgebraic geometric logic form complete lattices,
and we recover the similar result for $\Lambda$-bisimulations in
\cite[Proposition 3.7]{BakHan17} and \cite[Proposition 8.6]{BezGroVen19-report}.
\end{example}

\subsection{Characterisation via relation lifting}\label{subsec:rel-lift}

  Another property of
  bisimulations for $\cat{Set}$-coalgebras is that they can 
  be characterised via relation lifting (see e.g.~\cite[Section~2.2]{Sta11}),
  and that bisimilarity on a coalgebra $(X,\gamma)$ is a greatest fixpoint of a monotone operator
  on the lattice of relations $\Pow(X \times X)$.
  In this subsection and the following, we show that these results
  generalise to realm of $\rho$-bisimulations.
  
  %
  Given $X_1, X_2$ in $\cat{C}$, we shall define a monotone map  
  $$
    \fun{T}^{\rho} : \cat{Rel}(X_1, X_2) \to \cat{Rel}(\fun{T}X_1, \fun{T}X_2)
  $$
 which lifts $(B, \pi_1, \pi_2)$ in $\cat{Rel}(X_1, X_2)$ to
 $(\fun{T}^{\rho}B,\fun{T}^{\rho}\pi_1,\fun{T}^{\rho}\pi_2)$
  in $\cat{Rel}(\fun{T}X_1, \fun{T}X_2)$.
  In order to do so, consider the
  composition,
\begin{equation}\label{eq:sigma}
    \begin{tikzcd}[column sep=2em]
      \sigma_i : \fun{T}X_i \arrow[r, "\unitc_{\fun{T}X_i}"]
        & \fun{SPT}X_i \arrow[r, "\fun{S}\rho_{X_i}"]
        & [.5em] \fun{SLP}X_i \arrow[r, "\fun{SL}\ov{\pi}_i"]
        & [.5em] \fun{SL}\dpo{B},
    \end{tikzcd}
  \end{equation}
  for $i = 1, 2$.
For a concrete example of $\sigma_i$, see Example~\ref{exm:Trho-set} below.

\begin{definition}
  Given $\Bspan$
  in $\posRel$, 
  define
  $\fun{T}^{\rho}(B, \pi_1, \pi_2) = (\fun{T}^{\rho}B, \fun{T}^{\rho}\pi_1, \fun{T}^{\rho}\pi_2)$
  in $\cat{Rel}(\fun{T}X_1, \fun{T}X_2)$ as the pullback of
  \begin{tikzcd}[cramped, sep=1.5em]
    \fun{T}X_1 \arrow[r, "\sigma_1"]
      & \fun{SL}\dpo{B}
      & \fun{T}X_2 \arrow[l, swap, "\sigma_2"] 
  \end{tikzcd}.
\end{definition}
Observe that $(\fun{T}^{\rho}B, \fun{T}^{\rho}\pi_1, \fun{T}^{\rho}\pi_2)$
is a jointly mono span because it is a pullback.
Monotonicity of $\fun{T}^{\rho}$ follows from unravelling the definitions.
  We can now characterise $\rho$-bisimulations as in \cite{HerJac98}
  using the relation lifting $\fun{T}^\rho$.

\begin{theorem}\label{thm:beta-bis}
  A jointly mono span $(B, \pi_1, \pi_2)$ between two $\fun{T}$-coalgebras
  $(X_1, \gamma_1)$ and $(X_2, \gamma_2)$ is a $\rho$-bi\-sim\-ula\-tion
  if and only if there exists a morphism $\delta : B \to \fun{T}^{\rho}B$
  in $\cat{C}$ making diagram \eqref{eq:rho-bis-fill-in} commute.
  \begin{equation}\label{eq:rho-bis-fill-in}
      \begin{tikzcd}[row sep=1.5em]
        X_1   \arrow[d, "\gamma_1" swap]
          & B \arrow[l, "\pi_1" swap]
              \arrow[r, "\pi_2"]
              \arrow[d, dashed, "\delta"]
          & X_2
              \arrow[d, "\gamma_2"] \\
        \fun{T}X_1 
          & \fun{T}^{\rho}B
              \arrow[l, swap, "\fun{T}^{\rho}\pi_1"]
              \arrow[r, "\fun{T}^{\rho}\pi_2"]
          & \fun{T}X_2
      \end{tikzcd}
  \end{equation}
\end{theorem}
\begin{proof}   
  {\sl If $\delta$ exists, then $B$ is a $\rho$-bisimulation.} \;
  Suppose such a $\delta$ exists. In order to show that $B$ is
  a $\rho$-bisimulation, we need to show that the outer shell of the left diagram
  below commutes.
  Recall that $\unitc$ and $\unita$
  are the units of the dual adjunction
  $
  \begin{tikzcd}[cramped, sep=1.5em]
    \fun{P} : \cat{C}  \arrow[r,  shift left=1.7pt]
    & \cat{A} : \fun{S} \arrow[l, shift left=1.7pt]
  \end{tikzcd}
  $.
  \begin{equation}\label{prop-eq-beta-bis}
    \begin{tikzcd}[column sep=2.5em, row sep=1.7em]
        & \fun{L}\dpo{B}
            \arrow[dl, bend right=10, "\fun{L}\ov{\pi}_1" swap]
            \arrow[dr, bend left=10, "\fun{L}\ov{\pi}_2"]
            \arrow[d, "\unita_{\fun{L}\dpo{B}}"]
        &
        & [0em] \fun{L}\dpo{B}
            \arrow[r, "\unita_{\fun{L}\dpo{B}}"]
            \arrow[d, "\fun{L}\dpo{\pi}_i"]
        & \fun{PSL}\dpo{B}
            \arrow[d, "\fun{PSL}\dpo{\pi}_i" left]
            \arrow[ddd, bend left=30, shift left=10pt, "\fun{P}\sigma_i"] \\
      \fun{LP}X_1
            \arrow[d, "\rho_{X_1}" swap]
        & \fun{PSL}\dpo{B}
            \arrow[dl, bend right=5, "\fun{P}\sigma_1" {swap, pos=.2}]
            \arrow[dr, bend left=5, "\fun{P}\sigma_2" {pos=.3}]
        & \fun{LP}X_2
            \arrow[d, "\rho_{X_2}"] 
        & \fun{LP}X_i
            \arrow[r, "\unita_{\fun{LP}X_i}"]
            \arrow[d, "\rho_{X_i}"]
        & \fun{PSLP}X_i
            \arrow[d, "\fun{PS}\rho_{X_i}" left] \\
      \fun{PT}X_1
            \arrow[d, "\fun{P}\gamma_1" left]
        & \fun{PT}^{\rho}B
            \arrow[l, <-, "\fun{PT}^{\rho}\pi_1"]
            \arrow[r, <-, "\fun{PT}^{\rho}\pi_2" swap]
            \arrow[d, "\fun{P}\delta"]
        & \fun{PT}X_2
            \arrow[d, "\fun{P}\gamma_2" right]
        & \fun{PT}X_i
            \arrow[r, "\unita_{\fun{PT}X_i}"]
            \arrow[dr, equal, bend right=20]
        & \fun{PSPT}X_i
            \arrow[d, "\fun{P}\unitc_{\fun{T}X_i}" left] \\
      \fun{P}X_1
            \arrow[r, "\fun{P}\pi_1" swap]
        & \fun{P}B
        & \fun{P}X_2
            \arrow[l, "\fun{P}\pi_2"]
        &
        & \fun{PT}X_i
    \end{tikzcd}
  \end{equation}
  Commutativity of the bottom two squares follows from applying $\fun{P}$ to
  the diagram in \eqref{eq:rho-bis-fill-in}.
  The middle square commutes because of the definition of $\fun{T}^{\rho}B$.
  The top two squares commute because they are the outer shell of the 
  right diagram in \eqref{prop-eq-beta-bis}.
  In \eqref{prop-eq-beta-bis}, the right square commutes by definition of $\sigma_i$ (Equation~\ref{eq:sigma}).
  The other two squares commute by naturality of $\unita$
  and the lower triangle
  is a triangle identity of the dual adjunction.
  Therefore the outer shell commutes.
  
  \bigskip\noindent
  {\sl If $B$ is a $\rho$-bi\-sim\-ulation, then we can find $\delta$.} \;
  Suppose $(B, \pi_1, \pi_2)$ is a $\rho$-bi\-sim\-ulation. If we can prove that
  $
    \sigma_1 \circ \gamma_1 \circ \pi_1 = \sigma_2 \circ \gamma_2 \circ \pi_2
  $
  then we obtain $\delta$ as the 
  mediating map induced by the pullback which defines $\fun{T}^{\rho}B$,
  as shown below: 
  \begin{equation}\label{prop-bis-beta-2}
  \begin{tikzcd}[row sep=0em]
      & B \arrow[dl,  bend right=5, "\pi_1" {swap,pos=.4}]
          \arrow[dr, bend left=5, "\pi_2" {pos=.4}]
          \arrow[dd, dashed, "\delta"]
      & \\
    X_1   \arrow[dd, "\gamma_1" swap]
          &
      & X_2
          \arrow[dd, "\gamma_2"] \\
      & \fun{T}^{\rho}B
          \arrow[dl, bend right=3, "\fun{T}^{\rho}\pi_1" {swap, pos=.2}]
          \arrow[dr, bend left=3, "\fun{T}^{\rho}\pi_2" {pos=.3}]
      & \\
    \fun{T}X_1
          \arrow[dr, bend right=5, "\sigma_1" swap]
      &
      & \fun{T}X_2
          \arrow[dl, bend left=5, "\sigma_2"] \\
      & \fun{SL}Q
      &
  \end{tikzcd}
  \end{equation}
  
  We claim that the following diagram commutes. Since its outer shell is
  the same as the outer shell of \eqref{prop-bis-beta-2}, this proves the proposition.
  So consider:
  $$
  
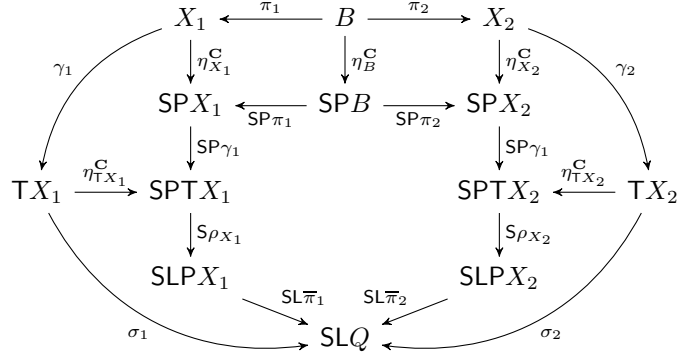
\begin{tikzcd}[row sep=1.7em]
      & X_1
          \arrow[d, "\unitc_{X_1}"]
          \arrow[ddl, bend right=30, "\gamma_1" swap]
      & B \arrow[l, "\pi_1" swap]
          \arrow[r, "\pi_2"]
          \arrow[d, "\unitc_B"]
      & X_2
          \arrow[d, "\unitc_{X_2}"]
          \arrow[ddr, bend left=30, "\gamma_2"]
      & \\
      & \fun{SP}X_1
          \arrow[d, "\fun{SP}\gamma_1"]
      & \fun{SP}B
          \arrow[l, "\fun{SP}\pi_1"]
          \arrow[r, "\fun{SP}\pi_2" swap]
      & \fun{SP}X_2
          \arrow[d, "\fun{SP}\gamma_1"]
      & \\
    \fun{T}X_1
          \arrow[r, "\unitc_{\fun{T}X_1}"]
          \arrow[ddrr, bend right=35, "\sigma_1" swap]
      & \fun{SPT}X_1
          \arrow[d, "\fun{S}\rho_{X_1}"]
      &
      & \fun{SPT}X_2
          \arrow[d, "\fun{S}\rho_{X_2}"]
      & \fun{T}X_2
          \arrow[l, "\unitc_{\fun{T}X_2}" swap]
          \arrow[ddll, bend left=35, "\sigma_2"] \\
      & \fun{SLP}X_1
          \arrow[dr, "\fun{SL}\ov{\pi}_1"]
      &
      & \fun{SLP}X_2
          \arrow[dl, "\fun{SL}\ov{\pi}_2" swap]
      & \\ [-1em] 
      &
      & \fun{SL}Q
      &
      &
  \end{tikzcd}
  $$
  Commutativity of the middle part follows from the fact that
  $B$ is a $\rho$-bisimulation. The four top squares commute because $\unitc$ is
  a natural transformation. The two remaining squares commute by definition of $\sigma_i$.
\end{proof}

We work out the explicit description of $\fun{T}^{\rho}$ in a special case:

\begin{example}\label{exm:Trho-set}
  Suppose we work with the classic dual adjunction
  $
  \begin{tikzcd}[cramped, sep=1.5em]
    \Pba : \cat{Set} \arrow[r, shift left=1.7pt]
      & \cat{BA} : \Uf \arrow[l, shift left=1.7pt]
  \end{tikzcd}
  $,
  $\fun{T}$ is an endofunctor on $\cat{Set}$, and the logic $(\fun{L}, \rho)$
  is given by predicate liftings and axioms (cf.~Example \ref{exm:pl}).
  Then the type of $\sigma_i$ is $\fun{T}X_i \to \Uf\fun{L}\dpo{B}$ and
  the ultrafilter $\sigma_i(t_i)$ 
  is determined by the elements of the form $\und{\lambda}(a_1, a_2)$ it contains,
  where $\lambda \in \Lambda$ and $(a_1, a_2) \in \dpo{B}$.
  Therefore the action of $\fun{T}^{\rho}$ on $(B, \pi_1, \pi_2)$
  is given by
  \begin{equation*}
    \begin{split}
      \fun{T}^{\rho}B = \{ (t_1, t_2) \in \fun{T}X_1 \times \fun{T}X_2 \mid
         \forall &\lambda \in \Lambda \text{ and $B$-coherent } (a_1, a_2) \\
         &\text{ we have } t_1 \in \lambda_{X_1}(a_1)
           \Leftrightarrow t_2 \in \lambda_{X_2}(a_2) \}.
    \end{split}
  \end{equation*}
  Informally, these are the pairs in $\fun{T}X_1 \times \fun{T}X_2$ 
  that cannot be distinguished by lifted $B$-coherent predicates. 
\end{example}


\subsection{Characterisation as a (post)fixpoint}\label{subsec:fix}

As for $\cat{Set}$-coalgebras, given a relation lifting of $\fun{T}$
and $\fun{T}$-coalgebras $(X_1,\gamma_1)$, $(X_2,\gamma_2)$,
we can define a map $\fun{T}^{\rho}_{\gamma_1, \gamma_2} \colon\cat{Rel}(X_1, X_2) \to \cat{Rel}(X_1, X_2)$ by,
essentially, taking inverse images under the $\gamma_i$.
This is a relational version of a predicate transformer on a coalgebra.

\begin{definition}\label{def:T-rho-gamma}
  Given $\fun{T}$-coalgebras $(X_1, \gamma_1)$ and $(X_2, \gamma_2)$
  and a jointly mono span $(B, \pi_1, \pi_2)$ between $X_1$ and $X_2$,
  define $\fun{T}^{\rho}_{\gamma_1, \gamma_2}(B, \pi_1, \pi_1) = 
  (\fun{T}^{\rho}_{\gamma_1, \gamma_2}B, \fun{T}^{\rho}_{\gamma_1, \gamma_2}\pi_1,
  \fun{T}^{\rho}_{\gamma_1, \gamma_2}\pi_2) \in \cat{Rel}(X_1, X_2)$ via the pullback
  $$
    \begin{tikzcd}[row sep=0em]
        &
        & \fun{T}_{\gamma_1,\gamma_2}^{\rho}B
            \arrow[dll, "\fun{T}_{\gamma_1,\gamma_2}^{\rho}\pi_1" swap]
            \arrow[drr, "\fun{T}_{\gamma_1,\gamma_2}^{\rho}\pi_1"]
        &
        & \\ [.5em]
      X_1   \arrow[dr, "\gamma_1" swap]
        &
        &
        &
        & X_2
            \arrow[dl, "\gamma_2"] \\ [-.5em]
        & \fun{T}X_1
            \arrow[dr, "\sigma_1" swap]
        &
        & \fun{T}X_2
            \arrow[dl, "\sigma_2"]
        & \\ [-.5em]
        &
        & \fun{SL}\dpo{B}
        &
        &
    \end{tikzcd}
  $$
  This is well defined because pullbacks are jointly mono spans.
\end{definition}

\begin{lemma}\label{lem:Trhogam-mon}
  The map $\fun{T}^{\rho}_{\gamma_1, \gamma_2}\colon \cat{Rel}(X_1, X_2) \to \cat{Rel}(X_1, X_2)$ is monotone. 
\end{lemma}
\begin{proof}
  If $(B, \pi_1, \pi_2) \leq (B', \pi_1', \pi_2')$ then there exists an
  $m : B \to B'$ such that $\pi_i = \pi_i' \circ m$.
  As a consequence the pullback $\dpo{B}'$ is a cone for $\dpo{B}$ and we have
  a mediating map $k : \dpo{B}' \to \dpo{B}$ satisfying $\dpo{\pi}_i' = \dpo{\pi}_i \circ k$.
  Unravelling the definitions reveals that $\fun{T}^{\rho}_{\gamma_1, \gamma_2}B$
  with its projections is a cone for $\fun{T}^{\rho}_{\gamma_1, \gamma_2}B'$,
  hence there is a (unique) map $t : \fun{T}^{\rho}_{\gamma_1, \gamma_2}B \to
  \fun{T}^{\rho}_{\gamma_1, \gamma_2}B'$ such that
  $\fun{T}^{\rho}_{\gamma_1, \gamma_2}\pi_i = \fun{T}^{\rho}_{\gamma_1, \gamma_2}\pi_i' \circ t$
  which witnesses that
  $\fun{T}^{\rho}_{\gamma_1, \gamma_2}(B, \pi_1, \pi_2) \leq
  \fun{T}^{\rho}_{\gamma_1, \gamma_2}(B', \pi_1', \pi_2')$.
\end{proof}

  As announced,
  $\rho$-bisimulations are precisely the post-fixpoints of
  $\fun{T}^{\rho}_{\gamma_1, \gamma_2}$.

\begin{theorem}\label{thm:rho-post-fix}
  A relation
  $
  \begin{tikzcd}[cramped, sep=1.5em]
    X_1 
      & B \arrow[l, swap, "\pi_1"] \arrow[r, "\pi_2"]
      & X_2
  \end{tikzcd}
  $
  is a $\rho$-bisimulation between $(X_1, \gamma_1)$
  and $(X_2, \gamma_2)$ if and only if
  $(B, \pi_1, \pi_2) \leq \fun{T}^{\rho}_{\gamma_1, \gamma_2}(B, \pi_1, \pi_2)$.
\end{theorem}
\begin{proof}
If $\Bspan$ is a $\rho$-bisimulation, then by  Theorem \ref{thm:beta-bis}
there is a map $\beta\colon B \to \fun{T}^\rho B$ such that
diagram \eqref{eq:rho-bis-fill-in} commutes. We then get a
map $\beta'\colon B \to \fun{T}^\rho_{\gamma_1, \gamma_2}B$
from the pullback property of $\fun{T}^\rho_{\gamma_1, \gamma_2}B$.
Conversely, given $\beta'\colon B \to \fun{T}^\rho_{\gamma_1, \gamma_2}B$,
we obtain  $\beta\colon B \to \fun{T}^\rho B$ from the pullback property of $\fun{T}^\rho B$.
\end{proof}

  Monotonicity of $\fun{T}^{\rho}_{\gamma_1, \gamma_2}$ and 
  the Knaster-Tarski fixpoint theorem imply:

\begin{corollary}
Under the assumptions of Proposition~\ref{prop:rel-jsl},
$\fun{T}^{\rho}_{\gamma_1, \gamma_2}$ has a greatest fixpoint,
and this greatest fixpoint is $\rho$-bisimilarity.
\end{corollary}

  This result encompasses a similar result in \cite[Proposition 6]{BakDitHan17}
  which states that $\Lambda$-bisimilarity for contingency logic
  is a bisimulation.

\begin{example}\label{exm:Trhogamma-set}
  We return to the classic setting of Example~\ref{exm:Trho-set}.
  Let $(B, \pi_1, \pi_2)$ be a relation 
  between $\fun{T}$-coalgebras $(X_1, \gamma)$ and $(X_2, \gamma_2)$.
  Then
  \begin{align*}
    \fun{T}^{\rho}_{\gamma_1, \gamma_2}B
      &= \{ (x_1, x_2) \in X_1 \times X_2 \mid 
            (\gamma_1(x_1), \gamma_2(x_2)) \in \fun{T}^{\rho}B \}.
  \end{align*}
  Informally, $\fun{T}^{\rho}_{\gamma_1, \gamma_2}B$
  consists of all pairs of worlds whose
  one-step behaviours are indistiguishable by lifted $B$-coherent predicates.
\end{example}

\section{Distinguishing power}
\label{sec:disting}

In this section
we compare the distinguishing power of $\rho$-bisimulations
with that of other semantic equivalence notions, and with logical equivalence.
We make the same assumptions here as at the start of Section~\ref{sec:rho-bis}.
Given a cospan  $(X_1,\gamma_1) \rightarrow (Y,\delta) \leftarrow (X_2,\gamma_2)$ in $\CoalgT$,
we call $(Y,\delta)$ a \emph{congruence} (of $\fun{T}$-coalgebras).

\subsection{Comparison with known equivalence notions}\label{subsec:other}

We briefly recall three coalgebraic equivalence notions, in descending order of
distinguishing power.
For more details, see e.g.~\cite[Definition~3.9]{BakHan17}.

\begin{definition}\label{def:equiv}
  Let $\Bspan$ be a jointly mono span between $(X_1,\gamma_1)$ and $(X_2,\gamma_2)$.
  Then $\Bspan$ is called a:
  \begin{enumerate}
    \item \emph{$\fun{T}$-bisimulation} if there is $t\colon B \to \fun{T}B$
          such that the $\pi_i$ become coalgebra morphisms;
    \item \emph{precocongruence}
          if its pushout $\po{\pi}_1: X_1 \to \po{B} \la X_2 : \po{\pi}_1$
          can be turned into a congruence between $(X_1,\gamma_1)$ and
          $(X_2,\gamma_2)$, more precisely, if there is
          $t\colon \po{B} \to \fun{T}\po{B}$ such that
          $$
            \begin{tikzcd}[row sep=0em]
                & B \arrow[dl, bend right=5, "\pi_1" {swap, pos=.35}]
                    \arrow[dr, bend left=5, "\pi_2"]
                & \\
              X_1   \arrow[dr, bend right=3, "\po{\pi}_1" {swap, pos=.65}]
                    \arrow[dd, "\gamma_1" swap]
                &
                & X_2   
                    \arrow[dl, bend left=3, "\po{\pi}_2"]
                    \arrow[dd, "\gamma_2"] \\ [-.3em]
                & \po{B}
                    \arrow[dd, dashed, "t"]
                & \\ [.3em]
              \fun{T}X_1
                    \arrow[dr, bend right=5, "\fun{T}\po{\pi}_1" {swap,pos=.6}]
                &
                & \fun{T}X_2
                    \arrow[dl, bend left=5, "\fun{T}\po{\pi}_2"] \\ [-.2em]
                & \fun{T}\po{B}
                &
            \end{tikzcd}
          $$
          commutes.
          $\po{\pi}_1$ and $\po{\pi}_2$ become coalgebra morphisms;
    \item \emph{behavioural equivalence} if it is a pullback in $\cat{C}$ of
          some cospan  $(X_1,\gamma_1) \rightarrow (Y,\delta) \leftarrow (X_2,\gamma_2)$
          in $\CoalgT$.
  \end{enumerate}
\end{definition}

  When $\fun{T}$ preserves weak pullbacks, all three notions coincide
  (when considering associated ``bisimilarity'' notions), 
  but in general, they may differ.
  In particular, expressive logics can generally only
  capture behavioural equivalence \cite{HanKupPac09}.
  The next proposition can be proved in the same way as \cite[Proposition~3.10]{BakHan17}.

\begin{proposition}
\label{it:T-bis} 
(i) Every $\fun{T}$-bisimulation is a $\rho$-bisimulation.
\label{it:preco}
(ii) Every precocongruence is a $\rho$-bisimulation.
\end{proposition}

The converse direction requires additional assumptions.

\begin{proposition}\label{prop:bis-preco}
  Suppose $\cat{C}$ has pushouts, $\fun{P}$ is faithful, and either
  \begin{itemize}
    \item[(i)] $\rho$ is pointwise epic; or
    \item[(ii)] $\mate{\rho}$ is pointwise monic and $\fun{T}$ preserves monos.
  \end{itemize}
  Then every $\rho$-bisimulation is a precocongruence.
  If, in addition, $\fun{T}$ preserves weak pullbacks, then
  $\rho$-bisimilarity coincides with 
  all three notions in Definition~\ref{def:equiv}. 
\end{proposition}
\begin{proof}
  Suppose
  $
  \begin{tikzcd}[sep=1.5em, cramped]
    X_1
      & B \arrow[l, "\pi_1" above]
          \arrow[r, "\pi_2" above]
      & X_2
  \end{tikzcd}
  $
  is a $\rho$-bisimulation with pushout $(\po{B}, \po{\pi}_1, \po{\pi}_2)$ be the pushout.
  We need to find a coalgebra structure $\zeta : \po{B} \to \fun{T}\po{B}$ which turns 
  $\po{\pi}_1$ and $\po{\pi}_2$ into coalgebra morphisms. It suffices to show that
  $
    \fun{T}\po{\pi}_1 \circ \gamma_1 \circ \pi_1 = \fun{T}\po{\pi}_2 \circ \gamma_2 \circ \pi_2,
  $
  because then the universal property of the pushout yields the desired $\zeta$.
  If $\fun{P}$ is faithful and $\rho$ is pointwise epic, then it suffices to 
  prove that
  $\fun{P}\pi_1 \circ \fun{P}\gamma_1 \circ \fun{PT}\po{\pi}_1 \circ \rho_{\po{B}}
   = \fun{P}\pi_2 \circ \fun{P}\gamma_2 \circ \fun{PT}\po{\pi}_2 \circ \rho_{\po{B}}$.
  This follows from the left diagram below, where the outer shell commutes
  because $(B, \pi_1, \pi_2)$ is a $\rho$-bisimulation and the top two squares
  commute by naturality of $\rho$.
  $$
  \begin{tikzcd}[row sep=0em]
      & \fun{LP}\po{B}
          \arrow[dl, bend right=5, "\fun{LP}\po{\pi}_1" {swap, pos=.25}]
          \arrow[dr, bend left=5, "\fun{LP}\po{\pi}_2" {pos=.25}]
          \arrow[dd, "\rho_{\po{B}}"]
      &
      & [1em]
      & \fun{SLP}\po{B}
      & \\
    \fun{LP}X_1
          \arrow[dd, "\rho_{X_1}" left]
      &
      & \fun{LP}X_2
          \arrow[dd, "\rho_{X_2}" right]
      & \fun{SLP}X_1
          \arrow[ru, bend left=5, "\fun{SLP}\po{\pi}_1" {pos=.8}]
      &
      & \fun{SLP}X_2
          \arrow[lu, bend right=5, "\fun{SLP}\po{\pi}_2" {swap,pos=.7}] \\
      & \fun{PT}\po{B}
          \arrow[dl, bend right=5, "\fun{PT}\po{\pi}_1"  {swap, pos=.25}]
          \arrow[dr, bend left=5, "\fun{PT}\po{\pi}_2" {pos=.25}]
      &
      &
      & \fun{T}\po{B}
          \arrow[uu, "\trans{\rho}_{\po{B}}"]
      & \\
    \fun{PT}X_1
          \arrow[dd, "\fun{P}\gamma_1" left]
      &
      & \fun{PT}X_2
          \arrow[dd, "\fun{P}\gamma_2" right]
      & \fun{T}X_1
          \arrow[ru, bend left=5, "\fun{T}\po{\pi}_1" {pos=.7}]
          \arrow[uu, "\trans{\rho}_{X_1}" left]
      &
      & \fun{T}X_2
          \arrow[lu, bend right=5, "\fun{T}\po{\pi}_2" {swap,pos=.6}]
          \arrow[uu, "\trans{\rho}_{X_2}" right] \\
      & \fun{P}\po{B}
          \arrow[dl, bend right=5, "\fun{P}\po{\pi}_1"  {swap, pos=.25}]
          \arrow[dr, bend left=5, "\fun{P}\po{\pi}_2" {pos=.25}]
      &
      &
      & \po{B}
      & \\
    \fun{P}X_1
          \arrow[dr, bend right=5, "\fun{P}\pi_1" swap]
      &
      & \fun{P}X_2
          \arrow[dl, bend left=5, "\fun{P}\pi_2"]
      & X_1
          \arrow[ru, bend left=5, "\po{\pi}_1" {pos=.7}]
          \arrow[uu, "\gamma_1" left]
      &
      & X_2
          \arrow[lu, bend right=5, "\po{\pi}_2" {swap,pos=.6}]
          \arrow[uu, "\gamma_2" right]
          \\
      & \fun{P}B
      &
      &
      & B
          \arrow[ul, bend left=5, "\pi_1"]      
          \arrow[ur, bend right=5, "\pi_2" swap]      & \\
  \end{tikzcd}
  $$
  
  Alternatively, suppose $\fun{P}$ is faithful
  (hence $\unitc\colon \Id_{\cat{C}} \to \fun{SP}$ is pointwise monic),
  $\rho^{\flat}$ is pointwise monic and $\fun{T}$ preserves monos.
  Then the transpose $\trans{\rho}_{\po{B}} : \fun{T}\po{B} \to \fun{SLP}\po{B}$
  of $\rho_{\po{B}}$ is monic, because
  $$
    \trans{\rho}_{\po{B}}
       = \fun{S}\rho_{\po{B}} \circ \unitc_{\fun{T}\po{B}}
       = \rho^{\flat}_{\fun{P}\po{B}} \circ \fun{T}\unitc_{\po{B}},
  $$
  so it suffices to show that
  $   \trans{\rho}_{\po{B}} 
                            \circ \fun{T}\po{\pi}_1 
                            \circ \gamma_1
                            \circ \pi_1
    = \trans{\rho}_{\po{B}} 
                            \circ \fun{T}\po{\pi}_2
                            \circ \gamma_2
                            \circ \pi_2.
  $
  But this follows from transposing the left diagram above, which yields
  the diagram on the right.

  When $\fun{T}$ preserves weak pullbacks, $\fun{T}$-bisimilarity coincides
  with behavioural equivalence \cite{Rut00}, and hence also with the largest
  precocongruence and $\rho$-bisimilarity.
\end{proof}

We note that condition (ii) in Proposition~\ref{prop:bis-preco}
entails that $(\fun{L},\rho)$ is expressive \cite[Thm.~4.2]{Kli07},
i.e., that logical equivalence implies behavioural equivalence.
In our abstract setting, \emph{logical equivalence} with respect to $(\fun{L},\rho)$ is
the kernel pair $(B,\pi,\pi')$ of the theory map $\th:X\to\fun{S}\Phi$.
Hence, $(\fun{L},\rho)$ is \emph{expressive} if $(B,\pi,\pi')$ is below
a behavioural equivalence in $\cat{Rel}(X,X)$.

\subsection{Hennessy-Milner type theorem}\label{subsec:hm}

  We now prove a partial converse to Proposition~\ref{prop:adequacy} (truth-preservation).
  We show that under certain conditions logical equivalence implies $\rho$-bisimilarity. 

\begin{theorem}\label{thm:ex}\label{thm:hm}
  Let 
  $
          \begin{tikzcd}[cramped, sep=1.2em]
            \cat{C'} \arrow[r, shift left=1.7pt]
              & \cat{A'} \arrow[l, shift left=1.7pt]
          \end{tikzcd}
          $
 be the dual equivalence induced by the dual adjunction 
 $
          \begin{tikzcd}[cramped, sep=1.2em]
            \cat{C} \arrow[r, shift left=1.7pt]
              & \cat{A} \arrow[l, shift left=1.7pt]
          \end{tikzcd}
$.
Suppose that
  \begin{itemize}
    \item $\cat{C}$ has $\RegEpiMono$-factorisations for morphisms with 
          domain $\in \cat{C'}$;
    \item $\cat{C'}$ is closed under regular epimorphic images;
    \item $\fun{S}$ is faithful and $\fun{L}$ preserves epis.
  \end{itemize}
  Then for all $\fun{T}$-coalgebras $(X, \gamma)$ with $X \in \cat{C'}$,
  logical equivalence, i.e., the 
  kernel pair $(B, \pi, \pi')$ of $\th_{\gamma} : X \to \fun{S}\Phi$, is a $\rho$-bisimulation.
\end{theorem}
\begin{proof}
  In order to prove that $(B, \pi, \pi')$ is a $\rho$-bisimulation, we need to show
  that the outer shell of
  \begin{equation}
    \begin{tikzcd}[row sep=1.3em]
        & \fun{L}\dpo{B}
            \arrow[dl, bend right=20, "\fun{L}\ov{\pi}" swap]
            \arrow[dr, bend left=20, "\fun{L}\ov{\pi}'"]
        & \\
      \fun{LP}X
            \arrow[dd, "\gamma^*" swap]
        & \fun{L}\Phi
            \arrow[l, "\fun{L}\llb \cdot \rrb_{\gamma}"]
            \arrow[r, "\fun{L}\llb \cdot \rrb_{\gamma}" swap]
            \arrow[d, "\alpha"]
            \arrow[u, dashed, "\fun{L}h"]
        & \fun{LP}X \arrow[dd, "\gamma^*"] \\
        & \Phi
            \arrow[dl, "\llb \cdot \rrb_{\gamma}" swap]
            \arrow[dr, "\llb \cdot \rrb_{\gamma}"]
            \arrow[d, dashed, "h"]
        & \\
      \fun{P}X
            \arrow[dr, bend right=15, "\fun{P}\pi" swap]
        & \dpo{B}
            \arrow[l, "\ov{\pi}"]
            \arrow[r, "\ov{\pi}'" swap]
        & \fun{P}X
            \arrow[dl, bend left=15, "{\fun{P}\pi'}"]\\ 
        & \fun{P}B
        &
    \end{tikzcd}
  \end{equation}
  commutes.

  From $B$ being the kernel pair of $\th_{\gamma}$ we have that
  $(\Phi, \llb \cdot \rrb_{\gamma}, \llb \cdot \rrb_{\gamma})$ is a cone
  for the pullback $\dpo{B}$. Hence we get a morphism
  $h : \Phi \to \dpo{B}$ such that the triangles left and right of $h$ commute,
  and it is easy to see that all the inner squares and triangles in the diagram
  on the right commute.
  Thus, in order to show that the outer shell commutes, it suffices to show that 
  $\fun{L}h$ is epic. 
By the assumption that $\fun{L}$ preserves epis,
it suffices to show that
$h : \Phi \to \dpo{B}$ is epic.
  Let $m \circ e$ be the $\RegEpiMono$-factorisation of $\th_{\gamma}$. Then
  the left diagram in \eqref{eq-t1-fact} commutes. 
  Since $m$ is monic the upper square is a pullback,
  and by Lemma \ref{lem-pb-po} 
  it is also a pushout.
  As a consequence, the lower square in the
  right diagram of \eqref{eq-t1-fact}, obtained from dualising the left one, is a pullback.
  \begin{equation}\label{eq-t1-fact}
    \begin{tikzcd}[row sep=0em, cramped]
        & B \arrow[dl, bend right=5, "\pi" swap]
            \arrow[dr, bend left=5, "\pi'"]
        & \\ [-0.3em]
      X     \arrow[dr, ->>, bend right=5, "e" {pos=.4}]
            \arrow[ddr, bend right=20, "\th_{\gamma}" swap]
        &
        & X \arrow[dl, ->>, bend left=5, "e" {swap,pos=.4}]
            \arrow[ddl, bend left=20, "\th_{\gamma}"] \\ [-.3em]
        & A \arrow[d, hook, "m"]
        & \\ [1.5em]
        & \fun{S}\Phi
        &
    \end{tikzcd}
    \qquad\quad
    \begin{tikzcd}[row sep=0, cramped]
        & \Phi
            \arrow[d, ->, "h"]
            \arrow[ddl, bend right=20, "\llb \cdot \rrb_{\gamma}" swap]
            \arrow[ddr, bend left=20, "\llb \cdot \rrb_{\gamma}"]
        & \\ [1.5em]
        & \fun{P}A
            \arrow[dl, hook, bend right=5, "\fun{P}e" {swap, pos=.4}]
            \arrow[dr, hook, bend left=5, "\fun{P}e" {pos=.4}]
        & \\ [-.3em]
      \fun{P}X
            \arrow[dr, hook, bend right=5, "\fun{P}\pi" swap]
        &
        & \fun{P}X
            \arrow[dl, hook, bend left=5,  "\fun{P}\pi'"] \\ [-.3em]
        & \fun{P}B
        &
    \end{tikzcd}
  \end{equation}
  Here $h$ denotes the adjoint transpose of $m$.
%
  Applying $\fun{S}$ to $h$ gives the morphism
  $\fun{S}h : \fun{SP}A \to \fun{S}\Phi$ which by assumption is isomorphic to
  $m$ (because $A \cong \fun{SP}A$). Since $\fun{S}$ 
  is faithful and $m$ is monic, $h$ and therefore $\fun{L}h$ are epic.
\end{proof}

\begin{example}\label{exm:classic-hm}
In the classic case, 
  $
    \begin{tikzcd}[cramped, sep=1.5em]
      \cat{Set}
            \arrow[r, shift left=1.7pt, ""]
        & \cat{BA}
            \arrow[l, shift left=1.7pt, ""]
    \end{tikzcd}
  $
  restricts to the full duality between finite sets and finite Boolean algebras.
$\cat{Set}$ has $\RegEpiMono$-factorisations \cite[Example~14.2(2)]{AdaHerStr90}.
In $\cat{Set}$ and $\cat{BA}$, all epis are regular and coincide with surjections \cite{AdaHerStr90,Ban10},
and finite sets are closed under surjective images.
The ultrafilter functor $\fun{S}$ is faithful.
If the logic functor $\fun{L}$ is given by predicate liftings and relations,
then by \cite[Remark 4.10]{KurRos12} it preserves regular epis,
and since all epis are regular, $\fun{L}$ preserves epis.
Applying Theorem~\ref{thm:hm}, we recover \cite[Theorem 4.5]{BakHan17}, and thereby all examples given there.
In particular, 
taking $(\fun{L},\rho)$ to be Hennessy-Milner logic
(Example~\ref{ex:LTS-HM-logic}), then
we recover from Theorem~\ref{thm:hm} that over finite labelled transition systems,
logical equivalence implies $\rho$-bisimilarity for Hennessy-Milner logic.
\end{example}

\begin{remark}\label{exm:posml--not-hm}
For positive modal logic from Examples~\ref{exm:pml-top} and \ref{exm:pml-top-bis},
we have not been able to show that the logic functor
$\fun{N} : \cat{DL} \to \cat{DL}$ preserves epis.
\end{remark}

\begin{example}\label{exm:hm-vec}
  We return to linear Hennessy-Milner logic from
  Examples~\ref{exm:mod-vec} and \ref{exm:mod-vec-bis}.
  The dual adjunction
  $
  \begin{tikzcd}[cramped, sep=1.2em]
     \cat{Vec}_{\Bbbk} \arrow[r, shift left=1.7pt]
      & \cat{Vec}_{\Bbbk} \arrow[l, shift left=1.7pt]
  \end{tikzcd}
  $
  restricts to the well-known self-duality of finite-dimensional vector spaces
  $\cat{FinVec}_{\Bbbk}$.
  The category $\cat{Vec}_{\Bbbk}$ has
  $\RegEpiMono$-factorisations \cite[Example~14.2]{AdaHerStr90}
  and the regular epis in both $\cat{Vec}_{\Bbbk}$ and $\cat{FinVec}_{\Bbbk}$
  are the surjections \cite[Example~7.72]{AdaHerStr90}.
  Moreover, the surjective image of a finite-dimensional vector space is 
  again fi\-nite-dimensional, and the functor $(-)^{\vee}$ is faithful. 
  Finally, since $\fun{L}$ is generated by
  predicate liftings and axioms it preserves surjections,
  so we can apply Theorem \ref{thm:hm} to conclude that
  logical equivalence and $\rho$-bisimilarity coincide
  on $\fun{W}$-coalgebras state-spaces in $\cat{FinVec}_{\Bbbk}$.
\end{example}

\begin{example}\label{exm:trace-not-hm}
  An example where logical equivalence does not imply $\rho$-bisimilarity
  is given by trace logic for labelled transitions systems (Example~\ref{ex:LTS-trace-logic}).    
  The conditions for Theorem~\ref{thm:hm} hold for trace logic, but
  the induced dual equivalence is in this case trivial, i.e., $\cat{C'}$ and $\cat{A}'$
  are the empty category, hence Theorem~\ref{thm:hm} does not tell us anything.
\end{example}

\subsection{Invariance under translations}\label{subsec:trans}

In this section we assume that $\cat{C}$ has pushouts.
  The example of Hennessy-Milner logic (Example~\ref{ex:LTS-HM-logic}) and
  trace logic (Examples~\ref{ex:LTS-trace-logic}~and~\ref{exm:trace-not-hm})
  is a situation where one logic is a reduct of the other.
  This can be considered a special case of translating a logic into another.
  We will show under which conditions $\rho$-bisimilarity is preserved under translations.
  To make this formal, we first generalise \cite[Definition 4.1]{KurLea12}.

\begin{definition}
Assume we are given a ``triangle situation'' as in diagram (\ref{eq:triangle}(a))
such that $\fun{P} = \fun{UP'}$,
and we have modal semantics $\rho' : \fun{L'P'} \to \fun{P'T}$ and $\rho : \fun{LP} \to \fun{PT}$.
  A \emph{translation} from $(\fun{L}',\rho')$ to $(\fun{L},\rho)$ is a natural transformation
  $\tau : \fun{LP} \to \fun{UL'P'}$ such that $\rho = \fun{U}\rho' \circ \tau$,
  see diagram (\ref{eq:triangle}(b)).
  \begin{equation}\label{eq:triangle}
    \begin{tikzcd}[row sep=.3em]
      & [1em]
        \cat{A'}
          \arrow[dd, shift left=3pt, bend left=9, "\fun{U}"]
          \arrow[dd, phantom, "\dashv"]
          \arrow[dd, <-, shift right=3pt, bend right=9, "\fun{F}" swap]
          \arrow[loop right, "\fun{L'}"]
      & [2em]
        \fun{LP}
          \arrow[dd, "\rho"]
          \arrow[r, "\tau"]
      & \fun{UL'P'}
          \arrow[dd, "\fun{U}\rho'"]
      & [.5em]
        \fun{FLP}
          \arrow[r, "\trans{\tau}"]
          \arrow[dd, "\fun{F}\rho" swap]
      & \fun{L'P'}
          \arrow[dd, "\rho'"]
    \\
    \cat{C}
          \arrow[ru, "\fun{P'}"]
          \arrow[rd, "\fun{P}" swap]
          \arrow[loop left, "\fun{T}"]
      &
      &
      &
      &
      & \\
      & \cat{A}
          \arrow[loop right, "\fun{L}"] 
      &     \fun{PT}
          \arrow[r, "="]
      & \fun{UP'T}
      &      \fun{FPT}
          \arrow[r, "\epsilon_{\fun{P}'\fun{T}}"]
      & \fun{P'T} \\ [.8em]
    {} \arrow[r, phantom, "\text{(a)}"]
      & {}
      & {} \arrow[r, phantom, "\text{(b)}"]
      & {}
      & {} \arrow[r, phantom, "\text{(c)}"]
      & {}
  \end{tikzcd}
  \end{equation}
  \end{definition}
\vspace{-.2em}\noindent
In (c), $\epsilon$ is the counit of $\fun{F} \dashv \fun{U}$
(which is adjoint to the identity) because $\fun{P} = \fun{UP'}$,
and $\trans{\tau}$ is the ($\fun{F} \dashv \fun{U}$)-adjoint of $\tau$.

\begin{proposition}\label{prop:trans-easy}
  Suppose $\tau$ is a translation from $\rho'$ to $\rho$
  and $(B, \pi_1, \pi_2)$ is a $\rho'$-bisimulation.
  Then it is also a $\rho$-bisimulation.
\end{proposition}
\begin{proof}
  Let $(\hat{B}, \hat{\pi}_1, \hat{\pi}_2)$ be the pushout of $(B, \pi_1, \pi_2)$.
  Since $B$ is assumed to be a $\rho'$-bisimulation the diagram on the left
  commutes.
  $$
    \begin{tikzcd}[row sep=0em, column sep=2.2em]
        &
        &
        & [2em]
        & \fun{LP}\hat{B}
            \arrow[dl, "\fun{LP}\hat{\pi}_1" swap]
            \arrow[dr, "\fun{LP}\hat{\pi}_2"]
            \arrow[dd, "\tau_{\hat{B}}"]
        & \\
        & \fun{L'P'}\hat{B}
            \arrow[dl, "\fun{L'P'}\hat{\pi}_1" swap]
            \arrow[dr, "\fun{L'P'}\hat{\pi}_2"]
        &
        & \fun{LP}X_1
            \arrow[dd, "\tau_{X_1}" swap]
            \arrow[dddd, bend right=60, shift right=5pt, "\rho_{X_1}" swap]
        &
        & \fun{LP}X_2
            \arrow[dd, "\tau_{X_2}"]
            \arrow[dddd, bend left=60, shift left=5pt, "\rho_{X_2}"] \\
      \fun{L'P'}X_1
            \arrow[dd, "\rho'_{X_1}" swap]
        &
        & \fun{L'P'}X_2
            \arrow[dd, "\rho'_{X_2}"]
        &
        & \fun{UL'P'}\hat{B}
            \arrow[dl, "\fun{UL'P'}\hat{\pi}_1" {swap, pos=.3}]
            \arrow[dr, "\fun{UL'P'}\hat{\pi}_2" pos=.3]
        & \\
        &
        &
        & \fun{UL'P'}X_1
            \arrow[dd, "\fun{U}\rho'_{X_1}"]
        &
        & \fun{UL'P'}X_2
            \arrow[dd, "\fun{U}\rho'_{X_2}"] \\
      \fun{P'T}X_1
            \arrow[dd, "\fun{P'}\gamma_1" swap]
        &
        & \fun{P'T}X_2
            \arrow[dd, "\fun{P'}\gamma_2"]
        &
        &
        & \\
        & \fun{P'}\hat{B}
            \arrow[dl, "\fun{P'}\hat{\pi}_1" swap]
            \arrow[dr, "\fun{P'}\hat{\pi}_2"]
        &
        & \fun{PT}X_1
            \arrow[dd, "\fun{P}\gamma_1" swap]
        &
        & \fun{PT}X_2
            \arrow[dd, "\fun{P}\gamma_2"] \\
      \fun{P'}X_1
            \arrow[dr, "\fun{P'}\pi_1" swap]
        &
        & \fun{P'}X_2
            \arrow[dl, "\fun{P'}\pi_2"]
        &
        & \fun{P}\hat{B}
            \arrow[dl, "\fun{P}\hat{\pi}_1" swap]
            \arrow[dr, "\fun{P}\hat{\pi}_2"]
        & \\
        & \fun{P'}B
        &
        & \fun{P}X_1
            \arrow[dr, "\fun{P}\pi_1" swap]
        &
        & \fun{P}X_2
            \arrow[dl, "\fun{P}\pi_2"] \\
        &
        &
        &
        & \fun{P}B
        &
    \end{tikzcd}
  $$
  Applying $\fun{U}$ to this diagram and putting the translation
  $\tau$ on top then yields the right diagram (using that $\fun{P} = \fun{UP'}$),
  and this proves that $(B, \pi_1, \pi_2)$ is a $\rho$-bisimulation.
\end{proof}

A sufficient condition for the converse is that the transpose $\trans{\tau}$ of $\tau$ is epic,
see diagram (\ref{eq:triangle}(c)).
Note that due to the adjunction $\fun{F} \dashv \fun{U}$, diagram (b) commutes if and only if (c) does.
Intuitively, $\trans{\tau} : \fun{FLP} \to \fun{L'P'}$ being epic formalises that
every modality in $\fun{L}'$ is a propositional combination of a modal formula of $\fun{L}$.
\begin{proposition}\label{prop:tau-flat}
  Suppose that $\trans{\tau}$ is pointwise epic. Then every $\rho$-bisimulation
  is a $\rho'$-bisimulation.
\end{proposition}
\begin{proof}
Commutativity of the outer shell of the following diagram will prove that
  $
  \begin{tikzcd}[cramped, sep=1.5em]
    X_1 
      & B \arrow[l, swap, "\pi_1"] \arrow[r, "\pi_2"]
      & X_2
  \end{tikzcd}
  $
  is a $\rho'$-bisimulation:
  $$
  
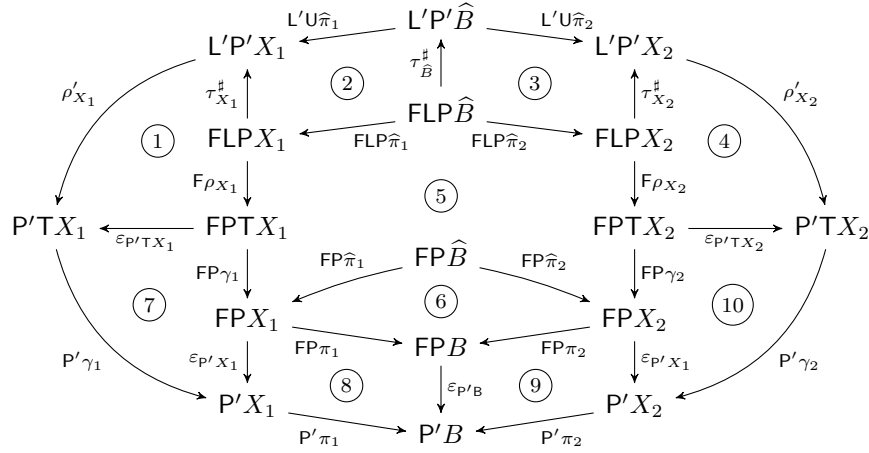
\begin{tikzcd}[row sep=.1em, column sep=large]
      &
      & \fun{L'P'}\po{B}
          \arrow[dl, "\fun{L'U}\po{\pi}_1" swap]
          \arrow[dr, "\fun{L'U}\po{\pi}_2"]
      &
      & \\ [-.6em]
      & \fun{L'P'}X_1
          \arrow[ddddl, bend right=30, "\rho'_{X_1}" swap]
          \arrow[ddddl, phantom, bend left=5, "\circled{1}"]
          \arrow[dr, phantom, "\circled{2}"]
      &
      & \fun{L'P'}X_2
          \arrow[ddddr, bend left=30, "\rho'_{X_2}"]
          \arrow[ddddr, phantom, bend right=5, "\circled{4}"]
          \arrow[dl, phantom, "\circled{3}"]
      & \\ [.6em]
      &
      & \fun{FLP}\po{B}
          \arrow[uu, "\trans{\tau}_{\po{B}}"]
          \arrow[dl, "\fun{FLP}\po{\pi}_1"]
          \arrow[dr, "\fun{FLP}\po{\pi}_2" swap]
          \arrow[dddd, phantom, "\circled{5}" {pos=.65}]
      &
      & \\ [-.6em]
      & \fun{FLP}X_1
          \arrow[uu, "\trans{\tau}_{X_1}"]
          \arrow[dd, "\fun{F}\rho_{X_1}" swap]
      &
      & \fun{FLP}X_2
          \arrow[uu, "\trans{\tau}_{X_2}" swap]
          \arrow[dd, "\fun{F}\rho_{X_2}"]
      & \\
      &
      &
      &
      & \\ [15pt]
    \fun{P'T}X_1
          \arrow[ddddr, bend right=30, "\fun{P'}\gamma_1" swap]
          \arrow[ddr, phantom, bend right=20, "{\circled{7}}" {pos=.65}]
      & \fun{FPT}X_1
          \arrow[dd, "\fun{FP}\gamma_1" swap]
          \arrow[l, "\epsilon_{\fun{P'T}X_1}"]
      &
      & \fun{FPT}X_2
          \arrow[dd, "\fun{FP}\gamma_2"]
          \arrow[r, "\epsilon_{\fun{P'T}X_2}" swap]
      & \fun{P'T}X_2
          \arrow[ddddl, bend left=30, "\fun{P'}\gamma_2"]
          \arrow[ddl, phantom, bend left=20, "{\circled{10}}" {pos=.65}] \\ [-.8em]
      &
      & \fun{FP}\po{B}
          \arrow[dl, bend right=5, "\fun{FP}\po{\pi}_1" {swap,pos=.25}]
          \arrow[dr, bend left=5, "\fun{FP}\po{\pi}_2" {pos=.3}]
          \arrow[dd, phantom, "\circled{6}"]
      &
      & \\ [.8em]
      & \fun{FP}X_1
          \arrow[dd, "\epsilon_{\fun{P'}X_1}" swap]
          \arrow[dr, "\fun{FP}\pi_1" swap]
      &
      & \fun{FP}X_2
          \arrow[dd, "\epsilon_{\fun{P'}X_1}"]
          \arrow[dl, "\fun{FP}\pi_2"]
      & \\ [-.6em]
      &
      & \fun{FP}B
          \arrow[dd, "\epsilon_{\fun{P'B}}"]
      &
      & \\ [.6em]
      & \fun{P'}X_1
          \arrow[dr, "\fun{P'}\pi_1" swap]
          \arrow[ur, phantom, bend right=6, "\circled{8}"]
      &
      & \fun{P'}X_2
          \arrow[dl, "\fun{P'}\pi_2"]
          \arrow[ul, phantom, bend left=6, "\circled{9}"]
      & \\ [-.6em]
      &
      & \fun{P'}B
      &
      &
  \end{tikzcd}
  $$
 %
  Cells 1 and 4 commute by diagram (c) in \eqref{eq:triangle}, 
  and cells 2 and 3 by naturality of $\trans{\tau}$.
  Commutativity of 5 and 6 together follows from applying $\fun{F}$ to the diagram
  witnessing the fact that
  $
  \begin{tikzcd}[cramped, sep=1.5em]
    X_1 
      & B \arrow[l, swap, ""] \arrow[r, ""]
      & X_2
  \end{tikzcd}
  $
  is a $\rho$-bisimulation.
  Commutativity of the remaining cells follows from the naturality
  of the counit $\epsilon$.
\end{proof}
  
  As a first example, we give a translation between Hennessy-Milner logic
  and trace logic.

\begin{example}\label{exle:triangle-hm-trace}
  Recall trace logic (Example~\ref{ex:LTS-trace-logic}) and Henessy-Milner
  logic (Example~\ref{ex:LTS-HM-logic}) for LTSs.
  Filling in the categories and functors in diagram (\ref{eq:triangle}(a)) we get:
  \begin{equation}\label{eq:triangle-hm-pos}
    \begin{tikzcd}[row sep=1em]
        & [1em]
          \cat{BA}
            \arrow[dd, shift left=3pt, bend left=9, "\fun{U}"]
            \arrow[loop right, "\Lhm"] 
            \arrow[dd, <-, shift right=3pt, bend right=9, "\fun{F}" swap]
            \arrow[dd, phantom, "\dashv"]
            \arrow[dl, bend right=11, shift right=0pt, "\Uf" {pos=.7}] \\
      \cat{Set}
            \arrow[ru, bend left=14, shift left=4pt, "\Pba"]
            \arrow[rd, bend right=11, shift right=0pt, "\cPow" {pos=.3}]
            \arrow[loop left, "\fun{T}"]
        &  \\
        & \cat{Set}
            \arrow[loop right, "\Ltr"]
            \arrow[lu, bend left=14, shift left=4pt, "\cPow"]
    \end{tikzcd} 
  \end{equation}
  We can define a translation $\tau : \Ltr \cPow \to \fun{U}\Lhm\Pow$
  by
  $$
    \tau_X
      : \Ltr \cPow X \to \fun{U}\Lhm\Pow X
      : \left\{ \begin{array}{rl}
          1      &\!\mapsto\, \top_{\Lhm\Pow X} \\
          (a, b) &\!\mapsto\, \langle a \rangle b
        \end{array}\right.
  $$
  Here $a \in A$, the set of labels, and $b \in \cPow X$.
  Then $\trans{\tau} : \fun{F}\Ltr\cPow X \to \Lhm\Pba X$ is surjective
  because each generator $\langle a \rangle b$ of $\Lhm\Pba X$
  is seen by some element in $\fun{F}\Ltr\cPow X$.
  Concretely, this is the case
  because formulae of Hennessy-Milner logic are precisely the Boolean
  combinations of trace logic formulae.
  Hence, in particular, $\trans{\tau}$ has surjective components,
  and in $\cat{BA}$ epis are the surjective Boolean homomorphisms.
  It now follows from Proposition~\ref{prop:tau-flat} that a
  $\rhotr$-bisimulation is a $\rhohm$-bisimulation (and the converse also holds).
\end{example}

In the setting of Examples~\ref{ex:LTS-HM-logic}, \ref{exm:pl} and \ref{exm:rho-bis-lambda},
where $\cat{A'}$ is a variety of algebras
and the logic $(\fun{L}, \rho)$ is given by predicate liftings and axioms,
we can consider the special case of \eqref{eq:triangle}
where $(\fun{L},\rho)$ is the ``modal reduct'' of $(\fun{L}', \rho')$.

\begin{example}\label{exm:translation-Var}
  Suppose $\cat{A}$ is a variety of algebras with
  free-forgetful adjunction $\fun{F} \dashv \fun{U}$. Let
  $(\fun{L}, \rho)$ be a logic for $\fun{T}$-coalgebras given by a collection $\Lambda$
  of predicate liftings and axioms (Example~\ref{exm:pl}).
  Then we can define $\fun{P_0} = \fun{U} \circ \fun{P}$,
  which has dual adjoint $\fun{S_0} = \fun{SF}$,
  where $\fun{S}$ is the dual adjoint of $\fun{P}$.
  Define the logic functor $\fun{L_0} : \cat{Set} \to \cat{Set}$ by
  $\fun{L_0}X = \{ \und{\lambda}_0(a_1, \ldots, a_n) \mid \lambda \in \Lambda, a_i \in X \}$
  and $\fun{L_0}f(\und{\lambda}_0(a_1, \ldots, a_n)) = \und{\lambda}_0(fa_1, \ldots, fa_n)$.
  \begin{equation}
      \begin{tikzcd}[row sep=.2em]
          & \cat{A}
              \arrow[dd, shift left=3pt, bend left=9, "\fun{U}"]
              \arrow[dd, phantom, "\dashv"]
              \arrow[dd, <-, shift right=3pt, bend right=9, "\fun{F}" swap]
              \arrow[loop right, "\fun{L}"] \\
        \cat{C}
              \arrow[ru, "\fun{P}"]
              \arrow[rd, "\fun{P_0}" {swap,pos=.6}]
              \arrow[loop left, "\fun{T}"]
          & \\
          & \cat{Set}
              \arrow[loop right, "\fun{L_0}"] 
      \end{tikzcd}
  \end{equation}
  
  Define $\tau : \fun{L_0P_0} \to \fun{ULP}$ by
  $\tau_X(\und{\lambda}_0(a_1, \ldots, a_n)) = \und{\lambda}(a_1, \ldots, a_n) \in \fun{ULP}X$.
  The logic $(\fun{L}, \rho)$ gives rise to the logic $(\fun{L_0}, \rho_0)$, where
  $\rho_0 = \fun{U}\rho \circ \tau : \fun{L_0P_0} \to \fun{P_0T}$.
  Then $\tau$ is a translation.
  One can verify that, in this situation, $\trans{\tau}$ is
  pointwise epic. Therefore a jointly mono span
  $
  \begin{tikzcd}[cramped, sep=1.5em]
    X_1 
      & B \arrow[l, swap, ""] \arrow[r, ""]
      & X_2
  \end{tikzcd}
  $
  in $\cat{C}$ is a $\rho$-bisimulation if and only if it is a $\rho_0$-bisimulation.
  Hence it suffices to look at the underlying sets when verifying
  whether a jointly mono span is a $\rho$-bisimulation.
\end{example}

\begin{example}\label{exm:trans-vec}
  If we apply the procedure from Example \ref{exm:translation-Var}
  to linear Hennessy-Milner logic (Example~\ref{exm:mod-vec}) we precisely get
  linear trace logic (Example~\ref{exm:trace-vec}).
  $$
    \begin{tikzcd}[row sep=.2em]
        & \cat{Vec}_{\Bbbk}
            \arrow[dd, shift left=3pt, bend left=9, "\fun{U}"]
            \arrow[dd, phantom, "\dashv"]
            \arrow[dd, <-, shift right=3pt, bend right=9, "\fun{F}" swap]
            \arrow[loop right, "\fun{L}"] \\
      \cat{Vec}_{\Bbbk}
            \arrow[ru, "(-)^{\vee}"]
            \arrow[rd, "(-)^{\circ}" {swap,pos=.6}]
            \arrow[loop left, "\fun{W}"]
        & \\
        & \cat{Set}
            \arrow[loop right, "\fun{L_0}"] 
      \end{tikzcd}
  $$
  Thus a span relation between two vector spaces
  (i.e., a linear subspace of $X_1 \times X_2$)
  is a linear trace logic bisimulation if and only if it is a
  linear Hennessy-Milner logic bisimulation.
  
  We can use this to transfer the Hennessy-Milner result from Example
  \ref{exm:hm-vec} for vector space logic with all vector space operators
  interpreted in finite linear weighted automata to the setting
  of linear trace logic.
  This relies on the fact that logical equivalence with respect to
  linear trace logic implies logical equivalence with respect to linear
  Hennessy-Milner logic. To see this, note that, using the axioms,
  we can rewrite any formula in linear HM logic to an equivalent formula
  of the form
  $$
    \phi ::= p \mid r \cdot \phi \mid \phi + \phi \mid \llangle a \rrangle p
  $$
  where $r \in \Bbbk$ and $\llangle a \rrangle$ is a finite sequence of the form
  $\langle a_1 \rangle\langle a_2 \rangle \cdots \langle a_n \rangle p$,
  with $a_i \in A$.
  Intuitively, this is the case
  because modalities are linear,
  and because we can view $0$ as shorthand for $0_{\Bbbk} \cdot p$.
  Now suppose two states $x, x'$ satisfy the same linear trace logic formulae,
  then we have $x \Vdash p$ iff $x' \Vdash p$ and
  $x \Vdash \llangle a \rrangle p$ iff $x' \Vdash \llangle a \rrangle p$.
  Since the interpretation of $r \cdot \phi$ and $\phi + \phi$ is computed
  pointwise, this implies that $x$ and $x'$ satisfy the same linear Hennessy-Milner
  logic formulae.
  
  As a consequence logical equivalence with respect to trace logic
  implies the existence of a bisimulation for linear Hennessy-Milner logic
  linking $x$ and $x'$, which in turn is also a linear trace logic bisimulation.
\end{example}

  Finally, we compare several logics that can be interpreted
  in topological spaces.

\begin{example}
  We squeeze the topological semantics for positive modal logic
  from Example \ref{exm:pml-top} between two
  other logics with varying base logics as in the following diagram:
  \begin{equation}
    \begin{tikzcd}[row sep=1.5em]
        & \cat{Frm}
            \arrow[d, "\fun{U'}"]
            \arrow[loop right, "\fun{N'}"] \\
      \cat{Top}
            \arrow[ru, "\fun{\Omega'}", bend left=22]
            \arrow[r, "\fun{\Omega}"]
            \arrow[rd, "\fun{\Omega_0}" swap, bend right=22]
            \arrow[loop left, "\fun{V}"]
        & \cat{DL}
            \arrow[d, "\fun{U}"]
            \arrow[loop right, "\fun{N}"] \\
        & \cat{Set}
            \arrow[loop right, "\fun{N_0}"]
    \end{tikzcd} 
  \end{equation}
  Here $\cat{Frm}$ is the category of frames and
  $\fun{\Omega}' : \cat{Top} \to \cat{Frm}$ is the functor that
  sends a topological space to its frame of opens. Let
  $\fun{N'} : \cat{Frm} \to \cat{Frm}$ be the functor given as in
  \cite[Section III4.3]{Joh82} (known also as the \emph{Vietoris locale})
  and define $\rho' : \fun{N'\Omega'} \to \fun{\Omega'V}$
  on generators by $\Box a \mapsto \dbox a$ and $\Diamond a \mapsto \ddiamond a$.
  The translation $\tau : \fun{N\Omega} \to \fun{U'N'\Omega'}$ given by
  $\Box a \mapsto \Box a$ and $\Diamond a \mapsto \Diamond a$
  is such that $\trans{\tau}$ is epic, 
  thus satisfies the assumptions of Proposition \ref{prop:tau-flat}.
  
  The bottom triangle is an instance of 
  Example~\ref{exm:translation-Var}. 
  We conclude that a jointly mono span
  $
  \begin{tikzcd}[cramped, sep=1.5em]
    X_1 
      & B \arrow[l, swap, ""] \arrow[r, ""]
      & X_2
  \end{tikzcd}
  $
  between $\fun{V}$-coalgebras $(X_1, \gamma_1)$ and $(X_2, \gamma_2)$ is a $\rho$-bisimulation
  if and only if it is a $\rho'$-bisimulation if and only if it
  is a $\rho_0$-bisimulation.
\end{example}

\section{Conclusion}
\label{sec:conc}

Our main question was whether we can characterise logical equivalence
for (possibly non-expressive) coalgebraic logics by a notion of bisimulation.
Towards this goal, we
generalised the logic-induced bisimulations in \cite{BakHan17}
for coalgebraic logics for $\cat{Set}$-coalgebras
to coalgebraic logics parameterised by a dual adjunction.
We identified sufficient conditions for when logical equivalence
coincides with logic-induced bisimilarity (Thm.~\ref{thm:ex}).
These are conditions on the categories in the dual adjunction,
and \emph{not} on the natural transformation $\rho$ defining (the semantics of)
the logic. In particular, we do not require the logic to be expressive.

We found that the distinguishing power of $\rho$-bisimulations depends on the modalities of the language but not on the propositional connectives. More generally, we showed that certain translations between logics preserve $\rho$-bisimilarity (Prop.~\ref{prop:tau-flat}).
Furthermore, as in the expressivity result of \cite{Kli07}, $\rho$-bisimilarity agrees with behavioural equivalence if the mate of $\rho$ is pointwise monic (Prop.~\ref{prop:bis-preco}). However, Example~\ref{exle:triangle-hm-trace} shows that this is not a necessary condition which raises the question whether one can characterise, purely in terms of $\rho$, when $\rho$-bisimilarity coincides with behavioural equivalence.

There are many other avenues for further research.
When is a congruence on complex algebras induced by a $\rho$-bisimulation?
Can we drop
in Theorem~\ref{thm:hm} the restriction to the subcategory if $\fun T$ is finitary?
Can we take quotients with respect to (the largest) $\rho$-bisimulation on a $\fun{T}$-coalgebra?

Moreover, the definition of $\rho$-bisimulation has a natural generalisation
to the order-enriched setting. This gives rise to \emph{$\rho$-simulations}. 
Can one prove an ordered Hennessy-Milner theorem where ``logical inequality'' is recognised by $\rho$-simulations?
Since this question naturally falls into the realm of order-enriched category theory, we will also seek a generalisation to the quantale-enriched setting, accounting for metric versions of simulation.

\bibliographystyle{plain}
\bibliography{bib-modal-bisim.bib}

\appendix
\section{Appendix}

\subsection{Some lemmas}

\begin{lemma}\label{lem-pb-split}
Pullbacks preserve split epimorphisms.
\end{lemma}
\begin{proof}
Suppose 
  $$
    \begin{tikzcd}[row sep=1.4em, cramped]
      B'     \arrow[rr, bend left=20, "\bar{g}"]
             \arrow[dr, bend right=10, "\id" swap] 
             \arrow[r, dashed, "g^*" swap]
        & A' \arrow[r, "v" swap] \arrow[d, "g"] & A \arrow[d, "f"] \\
        & B' \arrow[r, "w"] & B
    \end{tikzcd}
  $$
  is a pullback square and $f$ is split epic.
  Define $\bar{g} : B' \to A$ by $\bar{g} = f^*w$.
  Then we have $w = ff^*w = f\bar{g}$ hence a cone of the pullback.
  The fill-in $g^*$ satisfies $gg^* = \id_{B'}$, hence $g$ is split epic.
\end{proof}

\begin{lemma}\label{lem-pb-po}
  Let
  \begin{equation}\label{eq-sq-pb-po}
    \begin{tikzcd}[row sep=1.4em]
      A'    \arrow[r, "v"]
            \arrow[d, "f'"]
        & A \arrow[d, "f"] \\
      B'    \arrow[r, "w"]
        & B
    \end{tikzcd}
  \end{equation}
  be a pullback square such that $w, f$ are regular epic and
  $v, f'$ are split epic. Then the square is also a pushout.
\end{lemma}
\begin{proof}
  Note that every split epi is regular.
  Denote by $w_1, w_2$ the kernel pair of $w$
  (note that $w$ is the coequalizer of this pair) and similar for the other maps.
  Then we get the following diagram:
  $$
    \begin{tikzcd}[row sep=1.4em, cramped]
        & C'
            \arrow[r, dashed, "u"]
            \arrow[d, shift right=2pt, "f_1'" left]
            \arrow[d, shift left=2pt, "f_2'" right]
        & C \arrow[d, shift right=2pt, "f_1" left]
            \arrow[d, shift left=2pt, "f_2" right] \\
      A''   \arrow[d, dashed, "f''" left]
            \arrow[r, shift left=2pt, "v_1" above]
            \arrow[r, shift right=2pt, "v_2" below]
        & A'
            \arrow[d, "f'"]
            \arrow[r, "v"]
        & A \arrow[d, "f"] \\
      B''
            \arrow[r, shift left=2pt, "w_1" above]
            \arrow[r, shift right=2pt, "w_2" below]
        & B'
            \arrow[r, "w"]
        & B
    \end{tikzcd}
  $$
  The dashed arrow $f''$ can be obtained as the pullback of
  $ww_1 (= ww_2)$ and $vv_1 (= vv_2)$,
  and makes the horizontal coequalizers commute. (Similarly for $u$.)
  Moreover, $f''$ and $u$ are (split) epis by Lemma \ref{lem-pb-split}.

  Now let $h : B' \to D$ and $k : A \to D$ be such that $hf' = kv$.
  Then we have
  $$
    hw_1f'' = hf'v_1 = kvv_1 = kvv_2 = hf'v_2 = hw_2f''
  $$
  and since $f''$ is (split) epic it follows that $hw_1 = hw_2$.
  Since $w, w_1, w_2$ form a coequalizer, there exists $t : B \to D$
  such that $tw = h$.
  In a similar way we obtain $t' : B \to D$ such that $t'f = k$.
  Since $wf'$ is epic, it follows that $t = t'$.
  In general, unicity of $t$ follows from $wf'$ being epic.
  This proves that the square in \eqref{eq-sq-pb-po} is indeed a pushout square.
\end{proof}

\end{document}